\documentclass{article}
\usepackage{epsf}
\usepackage{emulateapj,pstricks}
\input{psfig}
\newbox\grsign \setbox\grsign=\hbox{$>$} \newdimen\grdimen \grdimen=\ht\grsign
\newbox\simlessbox \newbox\simgreatbox
\setbox\simgreatbox=\hbox{\raise.5ex\hbox{$>$}\llap
     {\lower.5ex\hbox{$\sim$}}}\ht1=\grdimen\dp1=0pt
\setbox\simlessbox=\hbox{\raise.5ex\hbox{$<$}\llap
     {\lower.5ex\hbox{$\sim$}}}\ht2=\grdimen\dp2=0pt

\def\lax    {{_<\atop^{\sim}}}

\def\lsi61  {{\hbox{LSI~+61~303}}}
\def\gx13   {{\hbox{GX13+1}}}

\def\Lo     {{L_{\odot}}}
\def\Mo     {{M_{\odot}}}

\def\etal   {{\it et~al.}}

\def\etal   {{\it et~al.}}

\def\doublespace {\smallskipamount=6pt plus2pt minus2pt
                  \medskipamount=12pt plus4pt minus4pt
                  \bigskipamount=24pt plus8pt minus8pt
                  \normalbaselineskip=24pt plus0pt minus0pt
                  \normallineskip=2pt
                  \normallineskiplimit=0pt
                  \jot=6pt
                  {\def\smallskip {\vskip\smallskipamount}}
                  {\def\medskip   {\vskip\medskipamount}}
                  {\def\bigskip   {\vskip\bigskipamount}}
                  {\setbox\strutbox=\hbox{\vrule 
                    height17.0pt depth7.0pt width 0pt}}
                  \parskip 12.0pt
                  \normalbaselines}
\def\onept5space {\smallskipamount=4pt plus2pt minus2pt
                  \medskipamount=8pt plus3pt minus3pt
                  \bigskipamount=16pt plus6pt minus6pt
                  \normalbaselineskip=16pt plus0pt minus0pt
                  \normallineskip=2pt
                  \normallineskiplimit=0pt
                  \jot=6pt
                  {\def\smallskip {\vskip\smallskipamount}}
                  {\def\medskip   {\vskip\medskipamount}}
                  {\def\bigskip   {\vskip\bigskipamount}}
                  {\setbox\strutbox=\hbox{\vrule 
                    height17.0pt depth7.0pt width 0pt}}
                  \parskip 12.0pt
                  \normalbaselines}
\def\myref#1  {\noindent \hangindent=24.0pt \hangafter=1 {#1} \par}
\def\bigref#1  {\noindent \hangindent=24.0pt \hangafter=2 {#1} \par}
\def\figs#1#2 {\item{#1} {#2} }

\usepackage{mymathptm}
\usepackage{flushrt}

\lefthead{David et al.}

\righthead{High Resolution Study of the Hydra A Cluster}

\slugcomment{accepted for publication in the  {\em The Astrophysical Journal}}
\def\ref#1{\noindent\hangindent=24.0pt\hangafter=1{#1}\par}
\def\la{\hbox{\rlap{$<$}\lower.5ex\hbox{$\sim$}\ }}
\def\ga{\hbox{\rlap{$>$}\lower.5ex\hbox{$\sim$}\ }}
\def\Mo{\rm{M_{\odot}}}

\begin{document}

\lefthead{David et al.}

\righthead{High Resolution Study of the Hydra A Cluster}

\slugcomment{accepted for publication in the  {\em The Astrophysical Journal}}

\title{A High Resolution Study of the Hydra A Cluster with Chandra:
Comparison of the Core Mass Distribution with Theoretical Predictions
and Evidence for Feedback in the Cooling Flow.}

\author{L.P. David,\altaffilmark{1} P.E.J. Nulsen,\altaffilmark{2} B.R. McNamara,\altaffilmark{1,3} W. Forman,\altaffilmark{1} C. Jones,\altaffilmark{1}T. Ponman,\altaffilmark{4},
B.~Robertson,\altaffilmark{5} and M.~Wise,\altaffilmark{6}}

\bigskip

\begin{abstract}
The cooling flow cluster Hydra A was observed during the orbital
activation and calibration phase of the Chandra Observatory.
While the X-ray image of the cluster exhibits complex structure
in the central region as reported in McNamara $etal$, the large scale
X-ray morphology of the cluster is fairly smooth.
A spectroscopic analysis of the ACIS data shows that the gas temperature
in Hydra A increases outward, reaches a maximum
temperature of 4~keV at 200 kpc, and then decreases slightly
at larger radii. The distribution of heavy elements is nonuniform,
with a factor of two increase in the Fe and Si abundances within
the central 100 kpc.  Beyond the central 100~kpc the Si-to-Fe abundance
ratio is twice solar, while the Si-to-Fe ratio of the central excess is
consistent with the solar value.
One of the more surprising results is the lack of
spectroscopic evidence for multiphase gas within the bulk of the cooling flow.
Beyond the central 30 kpc, the ACIS spectra are adequately fit with a single
temperature model. The addition of a cooling flow component does not
significantly improve the fit. Only within the central 30~kpc (where the cooling time
is less than 1~Gyr), is there spectroscopic evidence for multiphase gas.
However, the spectroscopic mass deposition rate is more than a factor of 10 less
than the morphologically derived mass accretion rate at 30~kpc.
We propose that the cooling flow region is convectively unstable due to heating
by the central radio source which significantly reduces the net accretion rate.
In addition, we show that the mass distribution within the
central 30-200~kpc region scales as $\rho_d \propto r^{-1.3}$,
intermediate between an
NFW and Moore profile, but with a best-fit NFW concentration parameter
($c_{NFW}=12$) approximately 3 times greater than that found
in numerical simulations. However, given the limited photon statistics,
we cannot rule out the presence of a flat-density core with
a core radius less than 30~kpc.

\end{abstract}

{\it Subject headings: cooling flows--galaxies:clusters:individual Hydra A}  

\vfill

\section{Introduction}

The launch of the Chandra X-ray Observatory (Weisskopf $\etal$ 2000) 
provides astronomers with the first 
opportunity to view clusters of galaxies with the same spatial 
resolution as ground based optical telescopes. This improved 
resolution is especially beneficial in the study of cooling 
flow clusters in which most of the interesting astrophysics is 
confined to the central 100~kpc (Fabian \& Nulsen 1977; 
Sarazin 1986; Fabian 1994).  
One of the great mysteries concerning cooling flows has always been 
the contrasting implications of the X-ray observations with observations
at other wavelengths. While X-ray observations suggest that a substantial 
amount of hot gas (up to 1000~$\Mo$~yr$^{-1}$) should be condensing out of 
the ambient medium within an extended region of radius 100~kpc,
activity at all other wavelengths is constrained to the central 
10~kpc (e.g., Hu $\etal$ 1985; Heckman $\etal$ 1989; 
McNamara \& O'Connell 1992; Donahue $\etal$ 2000). 
With Chandra's subarcsecond resolution, 
even scales of 10~kpc can be resolved in clusters at all redshifts.  

\smallskip

\footnotesize

\noindent
$^1$Harvard-Smithsonian Center for Astrophysics, 60 Garden St. Cambridge, MA 02138

\noindent
$^2$Department of Engineering Physics, University of Wollongong, Wollongong, NSW 2522, Australia

\noindent
$^3$Department of Physics and Astronomy,  Ohio University, Athens, OH 45701

\noindent
$^4$School of Physics \& Astronomy, University of Birmingham, Birmingham B15 2TT, UK

\noindent
$^5$University of Washington, Astronomy Department, Box 351580, Seattle WA 98195

\noindent
$^6$Massachusetts Institute of Technology, Center for Space Research, 70 Vassar Street, Building 37, Cambridge, MA 02139

\normalsize

Some of the first results from Chandra show that the central regions of
cooling flow clusters are not simple spherically symmetric systems, but are
morphologically and spectroscopically very complex.  Chandra has shown
that radio emission from the central galaxy in cooling flows can have a
significant impact on the X-ray morphology of the central regions.  X-ray
emission has been detected from the radio hotspots and the central point source
in 3C295 (Harris $\etal$ 2000; Wilson, Young \& Shopbell). Spectroscopic analysis of the Chandra data shows
that the emission from the hotspots is probably due to synchrotron
self-Compton and is not produced through an interaction between the
relativistic plasma in the radio lobes and the thermal gas.
The X-ray morphology of the central region of the Hydra A cluster 
(Abell 780) reveals
the presence of cavities in the X-ray emitting gas coincident with
the radio lobes (McNamara $\etal$ 2000). The Chandra A observation of Hydra A 
also
shows the presence of a heavily absorbed point source at the core
of the radio galaxy (McNamara $\etal$ 2000; Sambruna $\etal$ 2000).
The column density of absorbing material derived from
spectral analysis of the central point
source agrees with that observed in HI absorption (Taylor 1996).
The Chandra observation of the core of the Perseus cluster
shows a very complex X-ray morphology produced by interactions
between the hot gas and the radio emitting plasma and also low energy
absorption produced by an infalling gas rich galaxy (Fabian $\etal$ 2000).

A massive cooling flow was reported in Hydra A based on Einstein (David $\etal$ 1990),
ROSAT (Peres $\etal$ 1998), and ASCA (Ikebe $\etal$ 1997) observations with
estimated mass deposition rates ranging from $60$ to $600~\Mo$~yr$^{-1}$.
Einstein MPC and ASCA spectral analyses give an emission weighted temperature of
$\sim 4$~keV (David $\etal$ 1990; Peres $\etal$ 1998).
Temperature profiles have been derived from the ASCA data by Ikebe $\etal$ (1997),
Markevitch $\etal$ (1998), and White (2000).  The ASCA observation of Hydra A
also reveals the presence of an Fe abundance gradient (Ikebe $\etal$ 1997;
Finoguenov, David, \& Ponman 2000) with a radially increasing
silicon-to-iron ratio indicating a
predominance of Type II supernovae enriched material at large radii.
From the Chandra observations we are able to spectroscopically determine
the mass deposition rate, gas temperature, and abundance of heavy elements
at a much greater spatial resolution than previously possible.

In the first paper on Hydra A, we concentrated on the interaction
between the central FR Type I radio source (3C218)
and the ambient cluster
gas. In this paper we concentrate on the spectroscopic and
morphological properties of the cooling flow region,
derive high resolution temperature and abundance profiles of the
gas within the central 250~kpc, and determine the total gravitating
mass distribution in the core of the cluster. In addition,
we compare the observed mass distribution in Hydra A with theoretical
expectations from numerical simulations of dark matter halos
(e.g., Navarro, Frenk, \& White 1997; Moore $\etal$ 1998).

The remainder of this paper is organized as follows.  In $\S$~2
we discuss the large scale X-ray morphology of the Hydra A cluster,
the surface brightness profile, and the gas density distribution.  In $\S$~3 we
present the spectroscopic results concerning the
temperature and abundance profiles of the hot gas.  The total gravitating
mass and gas mass distributions are derived in $\S$~4 along with a
comparison with theoretical predictions.  The roles of weak and
strong shock heating by the central radio source in producing a convectively
unstable region within the cooling flow are discussed in
$\S$~5 and the main results are summarized
in $\S$~6.  We assume $\rm H_0$=70 km s$^{-1}$ Mpc$^{-1}$, $\Omega_{M}$=0.3,
and $\Omega_{\Lambda}$=0.7 throughout the paper.
At a redshift of 0.0538, the luminosity distance to Hydra A is
240~Mpc, and $1^{\prime\prime}=1.05$~kpc.

\section{X-Ray Morphology and Surface Brightness}

Hydra A was observed during the orbital activation and calibration
phase of the Chandra mission with both ACIS-I (October 30, 1999)
and ACIS-S (November 2, 1999). Due to flares in the background
charged particle flux, the data were screened to exclude all time intervals
when the background rate was more than than 20\% higher than the
quiescent rate. The resulting screened integration times are 24,100~s and
18,950~s for the ACIS-I and ACIS-S observations.  Only the standard grade
set (0,2,3,4,6) was used in our analysis. The gain
was corrected using the CIAO task $acis\_process\_events$ with
the most recent version of the gain table appropriate for observations
taken at a focal plane temperature of $-110^{\circ}$~C
(acisD1999-09-16gainN0004.fits).  This task also corrects for
charge transfer inefficiency (CTI)
effects in the front illuminated (FI) chips.

In McNamara $\etal$ (2000), we showed a high resolution image of the
combined ACIS-I and ACIS-S observations of Hydra A to illustrate the interaction between
the radio lobes and the ambient cluster gas.  To illustrate the
large scale X-ray morphology of Hydra A we use only
the ACIS-I observation due to the larger field of view and
lower background.  A background ACIS-I image was generated
by co-adding observations taken at a focal plane temperature
of $-110^{\circ}$~C in the Chandra public archive
after removing point sources from the observations.
The same gain corrections and background screening criteria were
applied to the observations resulting in a total integration
time of 73~ksec for the background image. The background image was
then mapped onto the sky
with the same aspect solution as in the Hydra A observation.  The
background subtracted image was corrected for detector and vignetting effects
using an exposure map generated by convolving the aspect histogram
with the mirror effective area and ACIS-I QE maps.   Finally,
the background-subtracted and exposure-corrected image was smoothed with a
$\sigma=4^{\prime\prime}$ Gaussian. Contours of the smoothed ACIS-I image
are shown in Figure 1 overlaid on the optical image of the cluster.  The
complex X-ray morphology within the central region discussed in
McNamara $\etal$ (2000) is evident in Figure 1.  Beyond the
central region, the X-ray morphology of Hydra A is fairly smooth with
slightly elliptical contours, indicating that the cluster has not
experienced a major merger in the recent past.  The X-ray isophotes are
aligned in the same SE-NW direction and have the same ellipticity as the
I-band isophotes shown in McNamara (1995). There is thus a remarkably
strong alignment between the stellar and gaseous components in the Hydra A
cluster from scales of a few kpc up to 300~kpc.

While the ACIS-I observation is useful for illustrating the large scale
2-D morphology of Hydra A, we use the ACIS-S observation
to derive an X-ray surface brightness profile due to the better
photon statistics on the back-illuminated (BI) S3 chip.
The BI chips on ACIS have higher QEs than the FI chips for photon
energies below 4~keV.  A broad band (0.3-7.0keV) surface brightness
profile derived from the S3 data is shown in Figure 2. 
A similar procedure was used to perform the background subtraction
as discussed above for the the ACIS-I data by co-adding 106 ksec
of ACIS-S3 data taken at a focal plane temperature of 
$-110^{\circ}$~C.  The same reductions were applied to this
background image as discussed above for the ACIS-I background
image.  Background count rates were then obtained from the same annuli 
as the total count rates.  The origin in the
surface brightness profile corresponds to the location of the central
point source.  The annuli were chosen to ensure
approximately 2,000 net counts per annulus.
Chandra's high resolution shows that there are two inflection points in the surface
brightness profile at approximately 100~kpc and 200~kpc.
It is obvious from Figure 2 that neither a single nor a double $\beta$ model
will provide a good fit to the data.  The break in the
surface brightness profile at 100~kpc is also evident in the ROSAT PSPC
observation discussed by Mohr, Mathiesen \& Evrard (1999).
Beyond the central 200~kpc, the surface brightness is approximately
characterized by a power-law with a slope corresponding to $\beta=0.7$, in good
agreement with the Einstein IPC results (David $\etal$ 1990) and
the ROSAT PSPC results (Mohr $\etal$ 1999; Vikhlinin, Forman, \& Jones 1999).

Due to the complexity of the surface brightness in the central
region of Hydra A, we use a deprojection technique to determine
the gas density profile rather than fitting a multivariate
function. 
The deprojection technique first determines the density of the gas in 
the largest annulus using the observed count rate and an assumed
emissivity and then determines the density at progressively smaller radii
after removing the projected emission from gas at larger radii.  
The fraction of the cluster emission in the outermost annulus
arising from gas at larger radii is corrected using the best fit
$\beta$ model to the PSPC observation of Hydra A in Mohr, Mathiesen
\& Evrard (1999).  We perform
the deprojection 100 times using a Monte Carlo technique, that 
randomizes the surface
brightness profile based on the observed errors.
The emissivity is assumed to be that of a 3.5~keV gas with a
40\% solar abundance of heavy elements.  In $\S 4$ below we show that the gas
temperature varies between 3.0 and 4.0~keV in this region and that the abundance
varies between 0.2 and 0.6 solar.  The assumption of a spatially independent
gas emissivity only introduces a 3\% uncertainty in the gas density.
The mean gas densities and rms deviations from the the 100 Monte Carlo
simulations are shown in Figure 3.  This figure shows that the density
profile gradually steepens from $\rho_{gas} \propto r^{-0.4}$
at 30~kpc to $\rho_{gas} \propto r^{-1.6}$ at 100~kpc, and there
is no obvious flat density core.

The isobaric cooling time, $t_c$, of the gas is shown in Figure 4.  Note that
$t_c$ is less than a Hubble time ($\sim 10^{10}$~yr) within the central
200~kpc, and less than 1~Gyr within the central 30~kpc. This inner radius is
important for the discussion below and appears to be the outer boundary
of an isentropic core of multiphase gas.

\section{Spectral Analysis}

The ACIS-S observation of Hydra A produced 160,000 source counts
within the central 250~kpc on the S3 chip.  Spectra were extracted in
8 concentric annuli centered on the point source with widths varying from 
$10^{\prime\prime}$ to $30^{\prime\prime}$ to ensure 
20,000 source counts per spectrum.
Spectra also were extracted with 40,000 net counts each, since
better photon statistics are required to determine the Si distribution.
We excised the emission from the central point source using 
a $2^{\prime\prime}$ radius circle. Background spectra were extracted from
the same regions in the background fits event list and were normalized by exposure time.
All spectra were extracted in PI space and rebinned to a minimum of 20 
counts per bin.  A single PI rmf file was generated
using mkrmf and the FEF file appropriate for the aim point on S3, node 1, at a 
focal plane temperature of $-110^{\circ}$~C. Area files were generated for 
each spectrum based on the mean off-axis angle of the extracted photons in each 
spectrum. We fitted each spectrum to a variety of XSPEC models:
an absorbed single temperature model, an absorbed single
temperature plus cooling flow model, an absorbed single temperature
plus intrinsically absorbed cooling flow model, and an absorbed two temperature 
model (see Tables 1-3).  Only data between 0.5-8.0~keV were included in the analysis. 
Below 0.5~keV the detector gain is nonlinear which is not yet properly treated
in the response matrices.  Above 8.0~keV, the charged particle background dominates
the source emission.

The ACIS-I data were extracted in a similar fashion, however, several modifications
are required to account for the nonuniformity of the FI chips due to radiation
induced CTI effects. Instead of a single rmf, a photon-weighted
rmf was generated for each spectrum. In addition, the CTI damage to the ACIS-I chips
produces a change in quantum efficiency at high energies (events far from the 
chip readout experienced significant grade migration, transforming acceptable grades 
to unacceptable grades) that we corrected using the ACIS-I QE maps at 
$\rm T = -110^{\circ}$~C.

The spectral fitting results for the inner three annuli of the ACIS-S data are
shown in Tables 1-3 and Figures 5-7.  
The spectrum from the inner $2^{\prime\prime}-17^{\prime\prime}$ is the most 
complex, and significant reduction in $\chi^2$ can be obtained by adding multiple
spectral components.  Table 1 shows that including a cooling 
flow component in the inner spectrum improves the fit with a reduction
in $\chi^2$ of 5 to 8 depending on whether $N_H$ is fixed to the 
galactic value or allowed to vary.  However, there is not an obvious improvement
when simply examining the best-fit of the two models to the data (see Figure 5).
The spectral analysis in the innermost annulus where the cooling time is
$\sim 10^8$~yrs, shows that less than 24\% (at the 90\% confidence limit)
of the total emission arises from gas condensing out of the flow. 
While the absorbed cooling flow model produces the minimum $\chi^2$ and largest
$\rm\dot M$, the best-fit galactic $N_H$ in this model is more than a factor of 
two below the Stark $\etal$ (1992) value.  
The intrinsic and galactic absorption are strongly correlated in 
the fitting process, and it is difficult to distinguish between them when 
fitting the spectrum of a low-redshift, low-$\rm\dot M$ cluster.
Fixing $N_H$ to the galactic value significantly reduces $\rm\dot M$ and 
limits the intrinsic absorption to $<3.2 \times 10^{20}$~cm$^{-2}$ (see Table 1).
Adding a cooling flow component to the second annulus only provides
a slight improvement to the fit (see Table 2), while adding a cooling flow component
to the third annulus does not produce any improvement (see Table 3). Figures
6 and 7 show how well the spectra from the second and third annuli
are fit with a single temperature model.  Beyond the central 30 kpc,
all of the spectra are well fit with single temperature models 
and the addition of more free parameters does not significantly improve the fits.

The lack of a strong spectroscopic signature for a massive cooling flow is 
not due to the small spatial binning of the spectra.  In Figure 8 we show a spectrum 
extracted from the central 100~kpc, excluding the central point source,  
fit to a single temperature plus cooling flow model with 
$\rm\dot M$ fixed at 140~$\Mo$~yr$^{-1}$ (the value obtained from a deprojection 
of the ROSAT PSPC data by Peres $\etal$ [1998] after adjusting to our
adopted cosmology). All other parameters
are treated as free parameters.  Figure 8 shows that there are very large
residuals below 2~keV and the resulting $\chi^2$/DOF = 1175/355.
Our results are in reasonable agreement with the spectral analysis of the ASCA
SIS data on Hydra A by Ikebe $\etal$ (1997) who obtained 
$\rm\dot M = 30 \pm 15$~$\Mo$~yr$^{-1}$
(scaled to our adopted cosmology).
The implications of these results for cooling flows will be discussed below.

\subsection{Temperature Profile}

The gas temperature profile derived from the ACIS-S and ACIS-I data is shown 
in Figure 9. The inner two temperature points for the ACIS-S data
were derived from the best-fit to a single absorbed temperature plus cooling flow model.
The remaining temperature points for the ACIS-S data were obtained 
without a cooling flow component in the spectral analyses. 
All ACIS-I spectra were fitted to single temperature models with 
$N_H$ fixed to the galactic value
and a variable abundance of heavy elements. Figure 9 
shows that there is good agreement between the two ACIS detectors
and that the gas temperature in Hydra A increases from 3
to 4~keV within the central 200~kpc as noted in McNamara $\etal$ (2000).  
It is very difficult to obtain any temperature estimates
from the ACIS-S data beyond 200~kpc due to the high background
in the S3 chip; however, 
the ACIS-I data show that the temperature drops to about 3~keV at 400-600~kpc .  
Previous temperature profiles for Hydra A were derived from ASCA data by
Ikebe $\etal$ (1997), Markevitch $\etal$ (1998), and White (2000).
The primary difference between the Chandra and ASCA profiles is the 
lower gas temperature in the center of the cluster and the steeper 
temperature gradient.  This is to be expected given the greater spatial resolution 
of Chandra.

\subsection{Abundance Gradients}

Intermediate temperature clusters like Hydra A emit strongly in 
both Fe L and Fe K lines.  To determine the Fe abundance of the gas,
we fitted the ACIS-S spectra in a broad energy band from 0.5-7.5~keV, that 
includes emission from both Fe L and K lines, and a hard band from 2.0-7.5~keV, 
that only includes Fe K lines.  The best-fit Fe abundances are shown in Figure
10 based on an absorbed VMEKAL model with $N_H$, all $\alpha$-processed 
elements, and Fe treated as free parameters.
This figure shows that there is excellent agreement between these two determinations
and that there is a strong Fe abundance gradient within the central 100~kpc.  
The agreement between the two Fe estimates shows that potential low
energy complications, e.g., excess absorption or cooling gas,
do not have a significant impact on the Fe abundance measurements.
An Fe abundance gradient in Hydra A also was reported
based on ASCA data by Ikebe $\etal$ (1997) and Finoguenov, David, \& Ponman (2000).
While the magnitude of the measured Fe gradient is
comparable between the ASCA and Chandra analysis, 
only Chandra has the spatial resolution to resolve the Fe distribution
within the central few arcminutes.  Figure 10 shows that the Fe gradient
is confined to the central 100kpc, while analysis of the ASCA 
data implied a flatter gradient extending to 200-300~kpc.

Significant constraints on the Si abundance can only be obtained from 
the set of spectra with 40,000 net counts. Figure 11 shows that the
Si distribution is similar to the Fe distribution with
a rapid decline within the central 100~kpc.  Beyond 100~kpc, the Si abundance
is essentially constant within the uncertainties. Due to the large errors on the
Si abundance, it is difficult to make conclusive comments on the 
radial variation in the Si/Fe abundance ratio.

\subsection{Deprojected Temperature and Entropy Profiles}

The projected temperature profile shows that there is a strong
positive gradient within the central 200~kpc (see Figure 9).  To determine
the deprojected temperature profile in this region, we first fitted the 
outer ACIS-S spectrum to a single MEKAL model, with $N_H$ and the abundance
of heavy elements treated as free parameters.  We then fitted the spectrum
extracted from the next smallest annulus to a two temperature model, with
the parameters of one of the models frozen to the best-fit values
derived from the outermost spectrum. The normalization of this
model is adjusted to account for the volume within
the outer shell projected along the line of sight toward
the next smallest shell.  This procedure
is continued inward, adding a MEKAL model for each successive spectrum,
with the free parameters of the remaining models frozen to the 
previously determined best-fit values.  Finally, the best-fit temperatures,
along with their associated errors, are fit to a power-law profile given 
by $T(r) = T_0 (r/10~\rm{kpc})^p$.  The best-fit is obtained with 
$T_0 = 2.73 \pm 0.07$~keV and $p=0.124 \pm 0.014$. The gas entropy 
is determined from the deprojected density profile and
the best-fit power-law temperature profile (see Figure 12). This
figure shows that the gas is approximately isentropic within the central 
30~kpc, and the gas entropy only increases significantly
at larger radii.

\section{Mass Distributions}

The gravitating mass distribution is determined from the equation
of hydrostatic equilibrium, given by:

$$M_{tot}(<r) = -{ {kTr} \over {\mu m_p G}} \left( {{d~ln~\rho_{gas}} \over {d~ln~r}} + {{d~ln~T} \over {d~ln~r}} \right) \eqno(1)$$

\noindent
We use a parametric form for the gas temperature (the best power-law
profile given above) along with the nonparametric density profile shown in Figure 3
to determine the mass distribution in Hydra A.  To reduce the noise in the 
$\rm{d~ln}~\rho_{gas} / \rm{d~ln}~r$ term, we determine the
gradient at $r_j$ by differencing the densities at $r_{j-2}$ and $r_{j+2}$.
The density gradients are assumed to be equal in the
innermost 3 points and the outermost 3 points.  
The cumulative gravitating mass, gas mass, Fe mass, and Si mass 
are shown in Figure 13.  In addition, we show the gravitational 
potential in Figure 14, with the potential normalized to zero at 250~kpc.  
Of the four mass components shown in Figure 13, the Fe and Si are the most centrally 
concentrated, the gas is the most extended, and the dark matter is intermediate 
between the two. 
The central concentration of a given mass component can be quantized
by calculating the ratio of the mass within 100 kpc to that within 200 kpc.
For the gas and dark matter this ratio is 27\% and 33\%, respectively.
Also shown in Figure 13 are the integrated Fe and Si masses 
that would be derived under the assumption of a uniform distribution of metals 
using the best-fit values derived from a spectral 
analysis of the integrated emission within the central 200~kpc.  
This figure shows that the assumption of a uniform distribution of heavy elements
underestimates the Fe and Si masses by a factor of two at small radii
and overestimates the Fe and Si masses by approximately 30-50\% at large radii. This 
has important consequences concerning the chemical enrichment history of clusters.

The variation in the gas mass fraction with radius is shown in Figure 15.
This figure shows that the gas mass fraction increases from 4\% at 10~kpc (several
times greater than a typical isolated giant elliptical)
up to 15\% at 200~kpc, which is typical of rich clusters 
(David, Jones, \& Forman 1995; White \& Fabian 1995; Evrard 1997).
The effects of the cluster environment on the X-ray characteristics of the central
dominant galaxy will be discussed in $\S~6$. 
The blue light within the central 47~kpc of Hydra A is 
$L_B = 9.2 \times 10^{10} \Lo$ (see $\S~6$) which gives $M/L_B= 76 \Mo / \Lo$.
Assuming $M/L_B =8 \Mo / \Lo$ for the stellar population of Hydra A gives
a stellar mass fraction of 10\% for a total baryon fraction of 15\%.  
The total baryon fraction within the central 47 kpc is thus comparable
to the gas mass fraction beyond 100 kpc.

\subsection{Comparison with Theoretical Expectations}

There has been a great deal of discussion in the literature concerning
the expected density profile of CDM halos and the observed density distribution of
galaxies and clusters.  Based on an extensive set of numerical simulations of 
the formation and evolution of dark matter halos, Navarro, Frenk \& White (1995,1998) 
concluded that dark matter halos have a universal density profile given by:

$${ \rho_d \over \rho_{crit}(z) }  = { \delta_c  \over { \left( { r \over r_s } \right)^{\gamma} \left( 1 + \left( {r \over r_s } \right)^{\alpha} \right)^{(\nu-\gamma)/ \alpha}}} \eqno(2)$$  

\noindent
where $\alpha=1$, $\nu=3$, $\gamma=1$, $\rho_{crit}(z)=3 H^2/8 \pi G$ 
is the critical density of the universe
at the cluster redshift, $r_s$ is a characteristic radius,
and $\delta_c$ is the central overdensity which can be expressed in terms of the 
concentration parameter ($c=r_{200}/r_s$; where $r_{200}$ is the radius within which 
the mean halo density is $200 \rho_{crit}$) as

$$\delta_c = { 200 \over 3 } { c^3 \over { \left[ \rm{ln}(1+c) -c/(1+c) \right]}}. \eqno(3)$$

\noindent
The NFW density profile varies from $\rho_d \propto r^{-1}$ at small radii
to $\rho_d \propto r^{-3}$ at large radii. 
For a given cosmology, the concentration parameter 
decreases with increasing halo mass.  Higher resolution simulations by 
Moore $\etal$ (1998) found a density profile similar to the NFW profile at 
large radii, but with a steeper
density profile at small radii ($\rho_d \propto r^{-1.5}$) corresponding
to $\alpha=1.5$, $\nu=3$, and $\gamma=1.5$.
Jing (2000) examined how the density profile depends on the dynamic
state of dark matter halos. He found significant variations in the density profile
of halos at the present epoch in his simulations.  By fitting the
simulated halo density profiles with an NFW profile, he found that the 
observed distribution in the concentration parameter can be 
described by a lognormal function.  Virialized systems have small
residuals when fit with the NFW profile and a mean concentration parameter
comparable to that predicted by NFW.  Unrelaxed systems produce
larger residuals and smaller concentration parameters when fit to
the NFW profile.

The primary difficulty encountered when these theoretical predictions
for CDM halos are compared to observations is the inferred presence of 
flat cores in the density profiles of dwarf and low surface brightness galaxies
(Flores \& Primack 1994; Burkert 1995; Moore $\etal$ 1998). For example,
Moore $\etal$ showed that the HI rotation curves of low surface brightness galaxies 
are best fit with a modified King profile ($\alpha=2$, $\nu=3$, and $\gamma=0$).
However, van den Bosch $\etal$ (2000) argue that the resolution of the
present HI data are insufficient to decouple the disk and halo components of
low surface brightness galaxies and cannot be used to place
significant constraints on the central density distribution.
They suggest that flat cores may only exist in low mass dwarf systems,
that are more susceptible to the effects of supernovae driven winds. 
A recent paper by Swaters, Madore, \& Trewhella (2000) demonstrates that
high resolution $\rm H_{\alpha}$ rotation curves of low surface 
brightness galaxies do indeed increase faster with radius than 
lower resolution HI rotation curves.
On cluster scales, Tyson, Kochanski, \& Dell'Antonio (1998)
analyzed strong lensing data on CL 0024+1654 and found evidence for a flat core, 
while Broadhurst $\etal$ (2000) concluded that a flat core is not
required by the strong lensing data for this cluster.
However, Shapiro, \& Iliev (2000) show that the mass distribution used
by Broadhurst $\etal$ predicts a galaxy velocity dispersion much 
greater than that observed, while the mass distribution derived by 
Tyson $\etal$ is consistent with the observed velocity dispersion.

The cumulative gravitating mass distribution of the Hydra A cluster in 
Figure 13 shows that there is little evidence for a flat
core with a core radius greater than approximately 30~kpc. 
We cannot exclude the presence of a smaller core, since we are only 
able to constrain the gas temperature within two annuli
in this region (see Figure 9).  Between
30 and 200~kpc, the integrated mass scales as $M(<r) \propto r^{1.7}$, or
equivalently, $\rho_d \propto r^{-1.3}$, which is slightly steeper than 
the NFW profile and slightly flatter than the Moore profile. The mass profile
is actually in very good agreement with simulations that incorporate
the effects of gas cooling and star formation by Lewis $\etal$ (2000), who
find $\rho_d \propto r^{-1.4}$.  Fitting the mass distribution in
Hydra A to these analytic models we obtain best-fit parameters of 
$\delta_c= (7.48 \pm 0.14) \times 10^4$ and $r_s=77 \pm 10$~kpc, 
$\delta_c= (4.02 \pm 0.36) \times 10^3$ and $r_s=234 \pm 25$~kpc, 
and $\delta_c= 4.54 \times 10^5$ and $r_s=26 \pm 2$~kpc, for the 
NFW, Moore, and modified King profiles, respectively. The best-fit
$\delta_c$ uniformly decreases with the level of mass concentration
in the models.  Eq. (3) is strictly valid only for the NFW profile and yields
a concentration parameter of $c_{NFW}=12.3 \pm 0.18$.
The best fit profiles for the three mass models are shown along with the
cumulative gravitating mass in Figure 16.  
This figure shows that both the NFW and Moore profiles 
provide reasonable fits to the data, supporting the results of numerical
CDM simulations.
Based on the rotation curve of the $H_{\alpha}$ disk in
Hydra A published in Melnick $\etal$ (1997), the
gravitating mass within 3.7~kpc is $3.9-7.8 \times 10^{10} \Mo$, for inclination
angles ranging from 0 to $45^{\circ}$.  The addition of this point
in Figure 16 favors the NFW profile.
The gravitational potential of the central galaxy should not affect our results 
since the gas temperature at 10~kpc is 2.7~keV. This 
is significantly hotter than the virial temperature of the central galaxy 
($kT \approx 0.6$~keV) derived from the stellar velocity dispersion of 
307~km~s$^{-1}$ (McElroy 1995).

The Hydra A cluster is approximately a $10M_*$ cluster. 
For the CDM$\Lambda$ cosmology used by NFW (essentially the same 
as that used to estimate distance dependent
quantities in this paper) the predicted concentration parameter for a 
$10M_*$ halo is 4.0, which is only one-third of the observed value. 
The high observed concentration of the Hydra A cluster 
implies an improbable collapse redshift
of $z \approx 4$ (using the expression given in the appendix of NFW).
The high concentration parameter in Hydra A cannot be resolved with NFW 
by appealing to the possible presence of residual substructure since
Jing (2000) finds that unrelaxed clusters yield 
concentration parameters lower than the NFW values.
Based on the NFW simulations, a concentration parameter of 
$c_{NFW} \approx 12$ is expected for a 0.01~$M_*$ dark matter halo,
which actually corresponds to the integrated mass within the central 10~kpc
of Hydra A.  In summary, we find that the total density distribution 
in the central region of the Hydra A cluster is fairly steep 
with $\rho_d \propto r^{-1.3}$.
The observed mass distribution is in reasonable agreement with CDM simulations,
but the derived concentration parameter is more representative of 
a galaxy-size halo rather than a rich cluster of galaxies. This analysis
will be extended to a sample of rich clusters in Robertson \& David (2000).

\section{Evidence for Feedback in the Cooling Flow}

The morphological mass accretion rate across a spherical shell j can be
estimated from:

$$\dot M_j = {L_j \over { \left( \Delta H_j + \Delta \phi_j \right) }} \eqno(4)$$

\noindent
where $L_j$ is the deprojected X-ray luminosity within shell j,
and $\Delta H_j$ and $\Delta \phi_j$ are the changes in 
gas enthalpy and gravitational potential across shell j.
Eq. (4) gives the mass accretion rate assuming steady-state and no 
mass deposition, in agreement with the spectroscopic results beyond
30 kpc.  This equation differs from that used in, e.g., Allen (2000)
because we do not include the X-ray emission arising from gas 
condensing out of the flow. 
Figure 17 shows the resulting morphological $\rm\dot M$ profile
using the deprojected X-ray luminosity, the best-fit power-law temperature 
profile, and the gravitational potential shown in Figure 14.  
Eq. (4) is only strictly valid beyond 30kpc and the 
only self-consistent solution is one with a constant 
$\dot M$, which is in fact the derived solution.
Beyond 30~kpc, the morphological $\rm\dot M$ is nearly constant at
300~$\Mo$~yr$^{-1}$ (see Figure 17).  This result provides strong
evidence that there is a steady, (nearly) homogeneous cooling flow beyond 
30~kpc.  Inside 30 kpc,
there is spectroscopic evidence for multiphase gas, but the 
spectroscopic mass deposition rate is more than a factor of 10 less 
than the morphological mass accretion rate at 30~kpc.  
Figure 17 also shows that the $\rm\dot M$ calculation is not very sensitive 
to the details of the gravitational potential.
The change in gas enthalpy within the central 200~kpc
is 2.5~keV per particle, while the change in gravitational potential
is only 0.5~keV per particle (see Figure 14).

The discrepancy between the morphological mass accretion rate at 30 kpc
and the spectroscopic $\rm\dot M$ within this radius
suggests that some heating mechanism 
may be suppressing the accretion of the cooling gas.  Many heating 
mechanisms have 
been proposed over the past two decades in the hopes of significantly reducing
the morphological mass accretion rate of cooling flow clusters.  
One of the most promising is heating from a central radio source, since 
approximately 70\% of cooling flow clusters have active central galaxies.  
Heinz, Reynolds, \& Begelman (1998) showed that shock heating by expanding 
relativistic plasma can induce observable X-ray features in the ambient 
cluster gas.  Tucker \& David (1997) showed that collective 
heating effects from relativistic plasma can generate a duty cycle between
accretion and outflow that can significantly reduce the integrated
mass accreted over the lifetime of a cluster. 
A similar model based on Compton heating by a central AGN in ellipticals 
has been developed by Ciotti \& Ostriker (1999).
Recently, Soker $\etal$ (2000) 
proposed a moderate cooling flow model due to heating by radio
sources that primarily affects the cooling rate of gas in the 
outer regions of cooling flows, significantly reducing their 
time-averaged mass accretion rates.  In addition to direct heating,
Churazov $\etal$ (2000) has shown that hot buoyant bubbles produced 
through shock heating by the expanding central radio source can 
dredge up cold material from the center of the cluster.
In this paper we further examine the role 
of weak and strong shock heating by a central radio source and discuss
the development of a convectively unstable region in cooling flows that 
can significantly reduce the net inflow of cooling gas.

\subsection{Heating by Weak Shocks}

The lack of evidence for gas deposition and the flattening of the
entropy profile near the center of the cluster suggest that the
cluster gas is being heated appreciably by the radio source.  
On the other hand, if this is the case, then the temperature decline towards
the center of the cluster is surprising, since it is an indication of
cooling.  If the age of this cluster since its last major merger is
about $10^{10}$ yr, then, in order to prevent the deposition of
$300~\rm{M_{\odot}}$~yr$^{-1}$ of cooled gas, the heating power needs to
make up for the radiative heat loss within the central 150~kpc
of the cluster, i.e.,  about $2.5\times10^{44}$~erg~s$^{-1}$.
This is plausible, since it is only 4 times the current radio
power (Ekers \& Simkin 1983).  Here we consider some of the details of
how heating by shocks from the radio source can prevent gas
deposition.

It is generally argued that the mechanical power of expanding radio
jets exceeds the radio power by a substantial factor.  This leads to
models in which the expanding radio jets drive a roughly spherical
shock into the surrounding gas (Heinz, Reynolds, \& Begelman 1998).
Shock heating due to such radio outbursts is a good candidate for the
required heating.  The typical lifetime of cluster center radio
sources is $10^7$ -- $10^8$ yr (e.g., Taylor $\etal$ 1990),
so that the high incidence of central radio sources in cooling flow
clusters (Burns $\etal$ 1997) requires that they undergo frequent
repeated outbursts.  We begin by considering the accumulated heating
effect of repeated weak shocks due to these outbursts.

The shock jump conditions may be written as

$$ {p_1 \over p_0} = 1 + {2\gamma\over \gamma + 1} y \eqno(5a)$$

\noindent
and

$${\rho_1 \over \rho_0} = {(\gamma + 1) (1 + y) \over \gamma + 1 + 
(\gamma - 1) y } \eqno(5b), $$ 

\noindent
where $p$ is the pressure, $\rho$ the density, $\gamma$ the ratio of
specific heats, and subscripts `0' and `1' refer to pre-shock and
post-shock conditions, respectively.  The quantity $y = \rho_0 v_{\rm
s}^2 / (\gamma p_0) - 1$ is the square of the shock Mach number minus
1, so that $y \ll 1$ for weak shocks and $y \gg 1$ for strong shocks.
The specific entropy jump for a weak shock is then

$$ \Delta S 
\simeq {2 \gamma k \over 3 (\gamma + 1)^2 \mu m_{\rm H}} y^3. \eqno(6)$$

\noindent
For a weak shock the kinetic energy of the shocked
gas is negligible compared to its increase in thermal energy.
Assuming that the postshock pressure excess is roughly uniform, we can
therefore estimate the shock strength by equating the mechanical
energy that drives the shock to the excess thermal energy within the
volume $V$ encompassed by the shock, i.e.,

$$ E_{\rm T} \simeq {1\over \gamma - 1} (p_1 - p_0) V. \eqno(7) $$

\noindent
Using the shock jump condition for $p_1$, this gives an expression for
$y$, which determines the specific entropy jump in the shock

$$ \Delta S \simeq {(\gamma - 1)^3 (\gamma + 1) k \over 12 \gamma^2
\mu m_{\rm H} } \left( E_{\rm T} \over p_0 V \right)^3. \eqno(8) $$

\noindent
The requirement that the shock is weak is $E_{\rm T} \ll p_0 V$.  If
the average repetition frequency of outbursts is $1/\tau$, then the
average heating rate per unit volume due to weak shocks is 

$$ H = \rho_0 T_0 {\Delta S \over \tau}
= {(\gamma - 1)^3 (\gamma + 1) p_0 \over 12 \gamma^2 \tau} \left(
E_{\rm T} \over p_0 V \right)^3. \eqno(9)$$

\noindent
Comparing this to the radiative cooling rate per unit volume, $R =
\rho_0^2 \Lambda(T_0)$, gives

$$ {H\over R} = {(\gamma - 1)^4 (\gamma + 1) \over 12 \gamma^2}
\left( t_{\rm c, 0} \over \tau \right) \left( E_{\rm T} \over p_0 V
\right)^3, \eqno(10)$$

\noindent
where the cooling time is $ t_{\rm c, 0} = p_0 / [(\gamma - 1) R]$.

Taking $\Lambda(T) \propto T^\alpha$, the ratio $H/R \propto
\rho_0^{-4} T_0^{-(2 + \alpha)} r^{-9}$, so that, if the gas is nearly
isothermal, $H/R$ increases with $r$ only if $\rho_0$ decreases more
rapidly than $r^{-9/4}$.  This is significantly steeper than the
observed $r$ dependence of $\rho$ in the region of interest, so that
$H/R$ is a decreasing function of $r$ within the central 150 kpc of
the Hydra A cluster.  Thus, heating by weak shocks can only dominate
cooling at small $r$.  The shock heating rate
would dominate the cooling rate everywhere inside the point where
they balance, i.e.,  $r=r_{\rm b}$, where $H=R$.  At small $r$ the net
heating time, $p_0 / [(\gamma - 1) (H - R)] \simeq p_0 / [(\gamma - 1)
H]$, would then be shorter than the cooling time and the gas is heated
rapidly.  Where the net heating is substantial it leads to convective
instability because the heating time is shortest at small $r$.  Where
the heating is not too fast (which includes the region where the
shocks are weak), convective instability drives mixing that will
result in a nearly isentropic core.

Details of the process leading to convection are complex, since as the
gas is heated it is redistributed and the pressure and density profiles
are flattened in order to maintain hydrostatic equilibrium.  At
$r_{\rm b}$ the entropy of the gas remains fixed.  To get a rough idea
of what part of the gas becomes isentropic we use the crude estimate

$$ t_{\rm S} = {S(r_{\rm b}) - S(r) \over {\left. dS/dt
\right|_r}} \eqno(11)$$ 

\noindent
for the time it takes for the entropy of the gas at $r$ to rise to the
entropy of gas at $r_{\rm b}$.  Assuming that the gas is (initially)
isothermal and $\rho_0(r) \propto r^{-\beta}$, the results above then
give 

$$ t_{\rm S} = {(\gamma - 1) \beta \ln (r_{\rm b} / r) \over (r_b /
r)^{9 - 4 \beta} - 1} t_{\rm c,0}(r). \eqno(12) $$ 

\noindent
Taking $\beta = 1.5$, roughly appropriate for the region outside the
isentropic core in Hydra, this gives $t_{\rm S} \simeq 0.1 t_{\rm c,
0}$ at $r = r_{\rm b}/2$.  Since the cooling time is already
relatively short in this region, this suggests that the bulk of the
gas within $r_{\rm b}$ should be isentropic. 

\subsection{Heating by Strong Shocks}

For the purpose of this discussion, heating by strong shocks adds one
significant complication, that a single strong shock can cause a
substantial entropy rise.  This can produce `bubbles' of gas with such
high entropy that they rise well outside the isentropic core
region ($r_{\rm b}$) before arriving at their equilibrium position.
Convection still creates an isentropic core within a few free-fall
times after passage of the shock, but some gas may rise well outside
this core.

We identify the roughly isentropic region at $r \lax 30$ kpc as
that where the average shock heating rate exceeds radiative cooling.
This determines the strength of shocks due to the radio outbursts as
follows.  
The shock repetition time, $\tau$, is highly uncertain, but, as
mentioned above, it needs to be comparable to the radio source
lifetime to account for the high incidence of central radio sources in
cooling flow clusters. 
Taking $\tau = 10^8$ yr, the cooling time at $r=30$ kpc is
too short relative to the shock repetition time for weak shocks to
balance the radiative cooling rate.  Instead we must appeal to
stronger shocks to provide the heating, but in that case balancing
the heating rate against the average cooling rate is complicated by
the large excursions in gas properties caused by the shocks.  Provided
that the gas is reasonably close to hydrostatic equilibrium most of
the time, we can use the ratio of the average rate of change of
entropy due to shocks to that due to radiative heat loss, i.e.,

$$ Q = {\Delta S / \tau \over R / (\rho T)}$$

\indent
$$= {t_{\rm c} \over \tau} \left[ \ln \left( 1 + {2 \gamma y \over
\gamma + 1} \right)+ \gamma \ln \left\{  \gamma + 1 + (\gamma - 1) y
\over (\gamma + 1) (1 + y) \right\}\right]. \eqno(13)$$

\noindent
This is essentially the same as the expression for $H/R$ above, except
that here $t_{\rm c}$ is an `average' cooling time, which we assume to
be close to the observed cooling time, and the expression for the
entropy jump in terms of $y$ is exact.  Taking $t_{\rm c} / \tau
\simeq 10$ at $r=30$ kpc and $\gamma=5/3$ then requires $y \simeq 2$ to
make $Q=1$ there.  If the observed pressure is close to the pre-shock
pressure, $p_0$, equation (7) still gives a reasonable estimate
for the shock energy.  Taking $y=2$, $kT = 3.3\rm\, keV$ and $n_{\rm
e} = 0.027\rm \, cm^{-3}$, as observed at $r=30$ kpc, we get $E_{\rm
T} \simeq 3.4 \times 10^{60}\rm\, erg$.  For $\tau=10^8\rm\, yr$, this
makes the average shock power about $1.1 \times 10^{45} \rm\, erg\,
s^{-1}$.  This is to be compared with the total power radiated by the
gas within $r=150\rm\, kpc$, i.e.,  about $2.5 \times 10^{44}\rm \, erg\,
s^{-1}$.  Given the large uncertainty in the shock energy, the
agreement is close, suggesting that the radiative cooling rate in the
central 150 kpc and shock heating due to radio outbursts are tied
together.  Note that these numbers agree for $\tau \simeq
4\times10^8\rm \, yr$. 

Direct shock heating does not fully explain the lack of gas
deposition, since the argument above suggests that it is insignificant
compared to radiative cooling outside the central 30 kpc.  In that
case, with no heat source to make up for radiative losses, cooling gas
from the region between 30 and 150 kpc flows into the central 30 kpc
at about $300 \rm\, M_\odot\, y^{-1}$.  If the system is to remain in
an approximately steady state, then an equivalent amount of gas must
be driven back into the region surrounding the core by each radio
outburst.  Again, taking $\tau \simeq 10^8\rm \, yr$, this means that
each radio outburst must drive about $3\times 10^{10}\rm \, M_\odot$
of gas from the core.  This is roughly 30\% of the mass of gas
in the 30 kpc core.  The core gas is roughly isentropic, so a strong
shock can certainly cause a substantial entropy inversion, i.e.,  gas
that can rise well outside the core.  It is not clear that as much as
30\% of the gas can be lifted from the core by a single shock.
More accurate numbers and more detailed modeling are needed to
determine if this is possible.

The detectability of the hot bubbles rising from the core is
determined by their size and speed.  They are unstable, and will tend
to break up and mix with ambient gas as they rise.  Given that they
are unlikely to rise faster than the thermal speed, i.e.,  about
$730\rm\, km\, s^{-1}$, they would take at least $1.3 \times 10^8$ yr
to reach 100 kpc.  This means that some hot bubbles should be present
most of the time during the cycles between outbursts.  The mass of
this hot gas is a few percent of the total gas mass within 100 kpc, so
it may be difficult to detect when it is well away from the core.  On
the other hand, it would be easily detected after the passage of the
shock, while it still forms a coherent, overpressured body.  Because
the hot gas rises relatively quickly (speeds of order the sound
speed), it fills a relatively small volume compared to inflowing
cooling gas (which is highly subsonic).  Thus, apart from some mixing,
the outflowing hot bubbles may have relatively little impact on the
steady inflow due to cooling.  

The shock itself would be easily detectable anywhere within the 30 kpc
core region (for $y=2$, the density jump is a factor of 2 and the
temperature jump a factor of 1.75).  At $r=30 \rm\, kpc$, the shock
speed is $v_{\rm s} = \sqrt{3 \gamma kT_0 / \mu m_{\rm H}} \simeq 1560
\rm\, km\, s^{-1}$, so that the shock takes about $10^7$ yr to reach
$r=30$ kpc.  This means that the shock is easily visible for at least
10\% of the time, or in at least 10\% of all comparable
sources.  If this model applies to many clusters, such shocks should
be detected soon.  If shocks are not detected, then shocks from the radio source cannot be
responsible for preventing the deposition of cooled gas and we would
need to find another heat source.

This description of gas flows around Hydra A presumes that the system
remains in a quasi-steady state.  Using the relationship between
stellar velocity dispersion and the mass of the central black hole (Gebhardt
$\etal$ 2000; Ferrarese \& Merrit 2000), the central galaxy in 
Hydra A should have a nuclear black hole
with a mass of $\simeq 5 \times 10^8 \rm\, M_\odot$.  As cooling gas accumulates
in the cluster core, the density of gas around this black hole
increases, presumably driving up the accretion rate until there is a
radio outburst.  The outburst heats and drives away the central gas,
thereby reducing the accretion rate and quenching the outburst.  This
provides a feedback mechanism connecting the average mechanical power
of radio outbursts to the total radiative power within the region
where there would otherwise be a cooling flow (i.e.,  where the cooling
time is shorter than the age of the system).  The critical condition
is that the net mass flow rate into the core region is kept small (the
spectroscopically determined mass deposition rate is less than
10\% of the rate at which cooling gas is delivered to the core in this
system).  In the absence of a detailed model for the outflow, it is
not clear that this requires very close balance between the average
mechanical heating rate and radiative cooling.  The mechanical heating
rate must at least match the cooling rate to prevent accumulation of
gas in the core, but any energy deposited outside the cooling flow
region has little impact on the current cooling rate and so is not
closely controlled by this feedback process.  Thus the mechanical
heating rate may exceed the radiative heat loss in the cooling flow
region.  In that case, the net heat input to the cluster could be
significant for the cluster as a whole.  For example, an average
excess heating rate of $10^{45}\rm\, erg\, s^{-1}$ for $10^{10}$ yr
deposited in $10^{14} \rm\, M_\odot$ of gas amounts to 1.0~keV per
particle.  This is close to what is required to account for
``preheating'' in clusters (e.g.,  Wu, Fabian, \& Nulsen 2000) and the
excess heating rate per unit mass could well be greater at early
times, when AGNs were more active.

In summary, for Hydra A, gas in the region from about 30 to 150 kpc is cooling
steadily and flowing inward, essentially as a (homogeneous) cooling
flow at a rate of about $300~\rm{M_{\odot}}$~yr$^{-1}$.  Repeated
outbursts from the central radio source drive shocks into the cluster
and heat the region $r < 30$ kpc strongly enough to prevent
all but about 10\% of the inflowing gas from cooling to low
temperatures.  Because it is heated from the center, the 30 kpc core
is convective and remains roughly isentropic.  During each outburst,
gas near the center of the core is heated so much that it rises
well outside the core before finding its equilibrium position.  On
average, the resulting mass outflow nearly balances the cooling inflow
to the core.  The rate of accretion onto the central black hole that
is the source of the radio outbursts is determined by the state of gas
in the core.  This link provides the feedback that keeps the net gas
flow rate into the core close to zero.  This model has much in common
with the model of Tabor \& Binney (1993) for gas flows in elliptical
galaxies.  Shock heating fits well with current models of radio
outbursts, but if shocks are the main heat source in the core, they
soon should be detected in some systems.  If shocks are not a major
heat source, then any other process by which the AGN heats the cluster
from its center would produce a similar general picture.  The major
factor distinguishing such models is the extent to which the heating
is intermittent.  Continuous heating would tend to make the whole of
the region, which would otherwise be depositing cool gas, isentropic.

\section{Source of the Central Abundance Gradient} 

Between 100 and 200 kpc the Fe and Si abundances are essentially constant 
with a Si/Fe abundance ratio twice solar (see Figures 10 and 11).  
This agrees with earlier results from ASCA observations of clusters 
indicating a significant role of Type II supernovae (SNe II) in the 
global enrichment of cluster gas (Mushotzky $\etal$ 1996, Fukazawa $\etal$ 1998). 
Within the central 100 kpc there is an excess of Si and Fe above
the globally averaged values.  Using an average Si abundance of 0.5 solar gives
an excess Si mass of $\Delta M_{Si} \simeq 1.4 \times 10^8 \Mo$ within 
the central 100 kpc.  Using an average Fe abundance of 0.25 solar gives
an excess Fe mass of $\Delta M_{Fe} \simeq 3.0 \times 10^8 \Mo$
within the same region.
The ratio $\Delta M_{Si} / \Delta M_{Fe}=0.5$ is essentially 
consistent with the solar ratio of 0.38.  Fukazawa $\etal$ (2000) 
and Irwin \& Bregman (2000) find that central metal excesses
are common in cD clusters with cooling flows based on ASCA and
BEPPOSAX observations.  Fukazawa $\etal$ also find that
the excess metals have a nearly solar abundance ratio. However, 
ASCA was unable to resolve 
the distribution of excess metals in clusters and accurately determine the Fe
and Si masses.  

The primary sources of metals in the central region 
of the Hydra A cluster are stellar winds and Type Ia Supernovae 
(SNe Ia) from the old stellar population in the central galaxy
and SNe II from the young stellar population in the blue
disk reported by McNamara (1995).
A central abundance excess will develop even if the gas 
is static due to the more extended distribution of the 
gas compared to the light from the central galaxy.  
Based on the spectroscopic mass deposition rate, only 10\% of the gas within 
the central 100 kpc should have condensed out of the hot gas over the 
cluster lifetime with the remainder being continuously enriched. 
Convection will flatten the abundance gradients to some extent but 
will not completely erase all gradients.  Convection will also dredge 
up material from within the central 10 kpc in the central galaxy
that should be enriched by stars 
with supersolar abundances, typical of the central regions of elliptical 
galaxies.

To determine the origin of the excess metals, we first compute the number of 
supernovae required to produce the excess Si and Fe masses. 
We adopt the supernovae yields given in Finoguenov, David, \& Ponman (2000),
that are averaged over a Salpeter IMF and given by 
$y_{Fe}(Ia) = 0.74 \Mo$, $y_{Fe}(II) = 0.070 \Mo$,
$y_{Si}(Ia) = 0.158 \Mo$, $y_{Si}(II) = 0.133 \Mo$.  
The excess Fe mass requires $4 \times 10^8$ SNe Ia.
Since the Si yield is nearly independent of supernova type, we 
can simply assume a yield of $0.14 \Mo$ per supernovae.  To produce the 
excess Si mass requires $10^9$ total supernovae, or $6 \times 10^8$ SNe II.
Assuming 1 SNe II for every $100 \Mo$ of star formation 
(appropriate for a Salpeter IMF) requires an average star formation rate of 
$60 \Mo$~yr$^{-1}$ over the past $10^{9}$~yr. 
McNamara (1995) showed that the observed luminosity and colors of 
the blue disk in Hydra A could only be reproduced by continuous star formation
over the past $10^{9}$~yr with star formation rates of 
of $\la 1 \Mo$~yr$^{-1}$.
Higher star formation rates are also permitted by the data, but only
when coupled with shorter periods of star formation.  It is thus unlikely that
the central metal excess in Hydra A was produced by star formation in the 
blue disk.

The total light of the central galaxy can be estimated from the 
photometric observations by Peterson (1986) who gives $M_V=-23.65$ 
and B-V=1.08 within a $1.65^{\prime}$ diameter aperture.
This gives $L_B=9.2 \times 10^{10} \Lo$ within the central 47 kpc 
radius using
our adopted cosmology.  To reproduce the observed excess Fe mass requires
an average SNe Ia rate of $4 \times 10^{-13} L_B^{-1}$~yr$^{-1}$ over
the past $10^{10}$ years, which is approximately twice 
Tammann's (1974) rate of $2.2 \times 10^{-13} L_B^{-1}$~yr$^{-1}$.
Tammann's rate is usually considered to be a factor of 4 too high 
(van den Bergh, McClure, \& Evans 1987), but our estimate does not 
include the contribution from stars beyond the central 50 kpc.
Given the large uncertainty in 
the present epoch SNe Ia rate in elliptical galaxies, 
the excess Fe can be accounted for by SNe Ia from
the old stellar population of the central galaxy.

Mass loss from stars as they evolve off the main sequence also can be 
a significant source of metals.  Using the Faber \& Gallagher (1976)
mass loss rate of $10^{-11} \Mo L_V^{-1}$~yr$^{-1}$  gives a 
stellar mass loss rate of $2.5 \Mo$~yr$^{-1}$ within the central 47 kpc.
If this gas has a solar abundance of heavy elements, and 
the present mass loss rate is assumed to hold over 
the lifetime of the galaxy, then stellar mass loss can only produce 
about 15\% of the observed Si and Fe excess, and only 
about 6\% of the total gas mass within this radius.

There is considerable evidence that the environment
of early-type galaxies has a significant impact on their X-ray properties
(e.g., Brown \& Bregman 1998; Helsdon $\etal$ 2000),
with isolated ellipticals having the lowest $L_x/L_B$ ratios and
the central dominant ellipticals in groups having the highest $L_x/L_B$ ratios.
Brighenti \& Mathews (1998) have suggested that much of the hot gas in 
ellipticals arises from the accretion of circumgalactic gas.
While the central galaxy in Hydra A is a rather extreme example,
since it is located at the center of a rich cluster, it is useful
to compare the gaseous properties within the central 10~kpc with that
found in samples of 
more isolated ellipticals.  The deprojected X-ray luminosity of the hot gas within the 
central 10~kpc is $L(0.5-2.0~\rm keV) = 8.6 \times 10^{42}$~ergs~s$^{-1}$ 
(excluding the central point source).  Compared with the 
samples of ellipticals in Brown \& Bregman and
Helsdon $\etal$, the central galaxy in Hydra A is 100 times overluminous
in X-rays for its observed $L_B$. 
The deprojected gas temperature within the central 10~kpc
is 2.7~keV which is approximately 3 times higher than a typical
elliptical. The higher temperature cannot result from heating by 
supernovae. Based on the calculations above, 
the production of the excess Si mass via SNe would only
heat the gas by 0.6~keV per particle.
Thus, the higher gas temperature must result from gravitational infall
and compression.  This is consistent with the result noted above
that the hot gas in the central
galaxy cannot be produced by stellar mass loss and must 
originate from the intracluster medium.

\section{SUMMARY}

We have presented a detailed analysis of the Chandra observations
of the Hydra A cluster.  The high spatial resolution of Chandra and 
the good photon statistics in these observations have allowed us to 
study the distribution of gas temperature, multiphase gas, and metals in the 
cluster core on scales of 10-20 kpc.  These high resolution
observations present significant challenges to the conventional cooling 
flow scenario and the processes by which the hot gas in clusters 
is enriched with heavy elements.

The large scale abundance ratio of heavy elements in the Hydra
A cluster determined from the Chandra observations indicates that the 
gas was initially enriched by a predominance of SNe II, in 
agreement with earlier ASCA observations. The uniform distribution 
of Fe and Si beyond the central 100 kpc suggests that this 
enrichment occurred prior to the last major merger, during which
any pre-existing gradients would have been erased.
Within the central 100 kpc, the observed excess of Fe and Si 
has a solar abundance ratio.

The ACIS-S spectra do not show any statistically significant evidence 
for multiphase gas beyond the central 30 kpc, even through the cooling 
time of the gas at this radius is only 1~Gyr.  Within 30 kpc, the 
spectroscopic mass deposition rate is roughly consistent with the observed star 
formation rate, but is more than an order of magnitude less than the 
morphological mass accretion rate at the same radius.  The gas is also 
isentropic within the central 30 kpc. Based on these observations,
we propose a scenario in which a small amount of the hot gas is able
to cool and accrete onto the central black hole in Hydra A and
trigger the formation of a radio jet which mechanically heats
the gas via strong shocks.  We show that the collective
effect of numerous weak shocks is energetically insufficient to have
a major impact on the cooling flow.  The boost in gas entropy after 
the passage of a strong shock makes the central gas convectively unstable 
and produces buoyant bubbles of hot gas.  Averaged over many
outbursts, the mass accretion rate in the outer cooling flow region
can be nearly balanced by the mass outflow rate in the expanding bubbles.
This scenario makes definitive predictions that can
be tested in the near future with Chandra and XMM.
About 10\% of cooling flow clusters should show evidence of strong shocks.
By the end of AO-2, Chandra will have observed approximately 100 groups and 
clusters so this prediction can be easily tested.  In addition, the greater
throughput of XMM-Newton may permit the 
spectroscopic detection of hot buoyant bubbles in the outer
parts of cooling flows. Based on our results, it is very likely
that some there will be some fundamental changes in the cooling flow
scenario over the next few years as data from 
Chandra and XMM-Newton continue to accumulate.

\acknowledgments

PEJN and TJP gratefully acknowledge the hospitality of the
Harvard-Smithsonian Center for Astrophysics.
We are grateful to J. Mohr for a very informative discussion
about recent theoretical and observational results concerning 
dark matter halos.  This work was partially supported by NASA contract 
NAS8-39073.  

\bigskip

\section*{References}
\ref{Brighenti, F. \& Mathews, W.G. 1998, ApJ, 495, 239.}

\ref{Broadhurst, T., Huang, X., Frye, B., \& Ellis, R. 2000, ApJ, 534, 15.}

\ref{Brown, B.A. \& Bregman, L.N. 1998, ApJ, 495, L75.}

\ref{Burkert, A. 1995, ApJ, 447, L25.}

\ref{David, L.P., Arnaud, K., Forman, W., \& Jones C. 1990, ApJ, 356, 32.}

\ref{Donahue, M., Mack, J., Voit, G.M., Sparks, W., Elston, R., \& Maloney, P.R. 2000, (astro-ph 0007062).}

\ref{Evrard, A.E. 1997, MNRAS, 292, 289.}

\ref{Faber, S.M. \& Gallagher, J.S. 1976, ApJ, 204, 365.}

\ref{Fabian, A.C. \& Nulsen P.E.J. 1977, MNRAS, 180, 479.}

\ref{Fabian, A.C. 1994, ARAA, 32, 277.}

\ref{Fabian, A.C., Sanders, J., Ettori, S., Taylor, G., Allen, S., Crawford, C., Iwasawa, K., Johnstone, R. \& Ogle, P. 2000, (astro-ph 0007456).}

\ref{Ferrarese, L. \& Merrit, D. 2000, ApJ, 539, L9.}

\ref{Finoguenov, A., David, L., Ponman, T. 2000 (preprint).}

\ref{Flores, R. \& Primack, J. 1994, ApJ, 427, L1. }

\ref{Fukazawa, Y., Makishima, K., Tamura, T., Ezawa, H., Xu, H., Ikebe, Y., Kikuchi, K., Ohashi, T. 1998, PASJ, 50, 187.}

\ref{Fukazawa, Y., Makishima, K., Tamura, T., Nakazawa, K., Ezawa, H., Ikebe, Y., Kikuchi, K., Ohashi, T. 2000, MNRAS, 313, 21.}

\ref{Gebhardt, K., Bender, R., Bower, G., Dressler, A., Faber, S., Filippenko, A., Green, R., Grillmair, C., Ho, L., Kormendy, J., Lauer, T., Magorrian, J., Pinkney, J., Richstone, D., \& Tremaine, S. 2000, ApJ, 539, L13.}

\ref{Heckman, T. M., Baum, S.A., van Breugel, \& W., McCarthy, P. 1989, Ap.J., 338, 48.}

\ref{Helsdon, S.F., Ponman, T.J., O'Sullivan, E., \& Forbes, D.A. 2000 (preprint).}

\ref{Heinz, S., Reynolds, C. \& Begelman, M. 1998, ApJ, 501, 126.}

\ref{Hu, E.M., Cowie, L.L., \& Wang, Z. 1985, Ap.J. Suppl., 59, 447.}

\ref{Ikebe, Y., Makishima, K., Ezawa, H., Fukazawa, Y., Hirayama, M., Honda H., Ishisake, Y., Kikuchi, K., Kubo, H., Murakami, T., Ohashi, T., Takahashi, T., \& Yamashita, K. 1997, ApJ, 481, 660.}

\ref{Irwin, J.A. \& Bregman, J.N. 2000 (astro-ph 0009237).}

\ref{Jing, Y. 2000, ApJ, 535, 30.}

\ref{Lewis, G., Babul, A., Katz, N., Quinn, T., Hernquist, L., \& Weinberg, D. 2000, ApJ, 536, 623}

\ref{Markevitch, M., Forman, W., Sarazin, C., \& Vikhlinin, A. 1998, ApJ, 503, 77.}

\ref{McElroy, D.B. 1995, A\&A Supp, 100, 105.}

\ref{McGaugh, S. \& de Block, W. 1998, ApJ, 499, 41.}

\ref{McNamara, B., \& O'Connell, R. 1992, ApJ., 393, 579.}

\ref{McNamara, B. 1995, ApJ, 443, 77.}

\ref{McNamara, B., Wise, M., Nulsen, P. E. J., David, L., Sarazin, C., Bautz, M., Markevitch, M.,Vikhlinin, A., Forman, W., Jones, C., Harris, D. 2000, ApJ, 534, 135L.}

\ref{Melnick, J., Gopal-Krishna, \& Terlevich, R. 1997, AA, 318, 337.}

\ref{Mohr, J., Mathiesen, B., \& Evrard. A. 1999, ApJ, 517, 627.}

\ref{Moore, B., Quinn, T., Governato, F., Stadel, J., \& Lake, G. 1999, MNRAS, 310, 1147.}

\ref{Mushotzky, R.F., Loewenstein, M., Arnaud, K.A., Tamura, T., Fukazawa, Y., Matsushita, K., Kikuchi, K., Hatsukade, I. 1996, ApJ, 466, 686.}

\ref{Navarro, J., Frenk, C., \& White, S. 1995, MNRAS, 275, 720.}

\ref{Navarro, J., Frenk, C., \& White, S. 1997, ApJ, 490, 493.}

\ref{Peres, C., Fabian, A., Edge, A., Allen, A., Johnstone, R., \& White, D. 1998, MNRAS, 298, 416.}

\ref{Peterson, C.J. 1986, PASP, 98, 486.}

\ref{Robertson, B. \& David, L. 2000 (in preparation).}

\ref{Sambruna, R., Chartas, G., Eracleous, M., \& Mushotzky, R. 2000, (astro-ph 0002201).}

\ref{Sarazin, C.S. 1986, Rev. Mod. Phys., 58, 1.}

\ref{Shapiro, P. \& Iliev, I. 2000 (astro-ph 0006353).}

\ref{Soker, N., White, R., David, L. \& McNamara, B 2000 (astro-ph 0009173).}

\ref{Swaters, R., Madore, B., \& Trewhella, M. 2000, ApJ, 531, L107.}

\ref{Tammann, G.A. 1974, Supernovae and Their Remnants, ed. C.B. Comsovici (Dordrecht:Reidel), p. 155.}

\ref{Taylor, G.B. 1996, ApJ, 470, 394.}

\ref{Thomas, P.A., Fabian, A.C., \& Nulsen, P.E.J.  1987, MNRAS, 228, 973.}

\ref{Tucker, W. \& David, L. 1997, ApJ, 484, 602.}

\ref{Tyson, J.A., Kochanski, G. \& Dell'Antonio, I. 1998, ApJ, 498, L107.}

\ref{van den Bergh, S., McClure, R.D., \& Evans, R. 1987, ApJ, 323, 44.}

\ref{van den Bosch, F., Robertson, B., Dalcanton, J. \& de Blok, W. 2000, AJ, 119, 1579.}

\ref{Vikhlinin, A., Forman, W., Jones, C. 1999, ApJ, 525, 47.}

\ref{Weisskopf, M., Tananbaum, H., Van Speybroeck, L., \& O'Dell, S. 2000, Proc SPIE 4012 (astro-ph 0004127).}

\ref{White, D. \& Fabian, A.C. 1995, MNRAS, 273, 72.} 

\ref{White, D. 2000, MNRAS, 312, 663.}

\ref{Wilson, A.S., Young, A.J. \& Shopbell, P.L. 2000, (astro-ph 0009308)}.

\newpage

\begin{table*}

\begin{center}

TABLE 1

Comparison of Spectral Fits for the $2^{\prime\prime}-17^{\prime\prime}$ annulus
\begin{tabular}{c|ccccccc}
\hline\hline
Model & kT & $N_H$ & Z & $\rm\dot M$ & $N_{Hi}$ & $kT_{low}$ & $\chi^2$/DOF \\
& (keV) & $(10^{20} \rm{cm}^{-2})$ & (solar) & $(\Mo \rm{yr}^{-1})$ & $(10^{20} \rm{cm}^{-2})$ & (keV)  &  \\
\hline
WABS*MEKAL & $2.78^{+.08}_{-.09}$ & 4.88 & $0.36^{+.06}_{-.05}$ & - & - & - &  250/181 \\
 & $2.88^{+.18}_{-.11}$ & $3.65^{+.79}_{-.80}$ & $0.41^{+.07}_{-.06}$ & - & - & - & 242/180 \\
&&&&&&& \\
WABS*(MEKAL& $3.00^{+.15}_{-.13}$ & 4.88 & $0.44^{+.06}_{-.06}$ & $8.9^{+4.1}_{-4.9}$ & - & - & 242/180 \\
+ MKCFLOW) & $3.11^{+.14}_{-.15}$ & $3.63^{+.57}_{-.68}$ & $0.49^{+.07}_{-.07}$ & $6.3^{+4.3}_{-4.3}$ & - & - & 237/179 \\
&&&&&&& \\
WABS*(MEKAL& $3.05^{+.11}_{-.17}$ & 4.88 & $0.44^{+.06}_{-.06}$ & $8.7^{+4.3}_{-3.9}$ & $<3.2$ & - & 241/179 \\
+ZWABS*MKCFLOW) & $3.10^{+.14}_{-.15}$ & $<2.3$ & $0.48^{+.08}_{-.06}$ & $29^{+12}_{-17}$ & $38.0^{+4.3}_{-3.9}$ & - & 232/178 \\
&&&&&&& \\
WABS*(MEKAL& $4.08^{+3.9}_{-1.02}$ & 4.88 & $0.31^{+.06}_{-.05}$ & - & - & $1.93^{+.68}_{-.32}$ & 240/179 \\
+MEKAL) & $3.72^{+2.4}_{-.90}$  & $3.56^{+.80}_{-0.5}$ & $0.37^{+.06}_{-.07}$ & - & - & $1.91^{+.59}_{-.72}$ & 233/178 \\
\hline
\end{tabular}
\end{center}
Notes: The results of fitting the ACIS-S data to a variety of spectral models (defined in
column 1). $kT$ is the gas temperature in single temperature fits and
the highest temperature in two-temperature fits, $N_H$ is the hydrogen column density,
$Z$ is abundance of heavy elements relative to the solar value, $\dot M$ is the
mass deposition rate in models with a cooling flow component, and $kT_{low}$ is the
lower gas temperature in the two-temperature fit.
When fitting the single temperature plus cooling flow model, the abundance
of heavy elements is linked between the two components and the lowest
gas temperature is fixed at 0.02~keV.  Two rows are shown for each model.
The top row shows the results with $N_H$ fixed to the galactic value and the
second row shows with results with $N_H$ free.  All error bars are shown at
the 90\% confidence level for one interesting parameter ($\chi^2_{min}+2.71$).

\end{table*}

\begin{table*}

\begin{center}

TABLE 2

Comparison of Spectral Fits for the $17^{\prime\prime}-29^{\prime\prime}$ annulus
\begin{tabular}{c|cccccc}
\hline\hline
Model & kT & $N_H$ & Z & $\rm\dot M$ & $N_{Hi}$ &  $\chi^2$/DOF \\
& (keV) & $(10^{20} \rm{cm}^{-2})$ & (solar) & $(\Mo \rm{yr}^{-1})$ & $(10^{20} \rm{cm}^{-2})$ &  \\
\hline
WABS*MEKAL & $2.97^{+.11}_{-.13}$ & 4.88 & $0.35^{+.06}_{-.05}$ & - & - & 249/190 \\
 & $3.15^{+.14}_{-.15}$ & $3.44^{+.66}_{-.67}$ & $0.41^{+.06}_{-.06}$ & - & - & 236/189 \\
&&&&&& \\
WABS*(MEKAL& $3.15^{+.17}_{-.17}$ & 4.88 & $0.38^{+.06}_{-.05}$ & $7.6^{+4.3}_{-4.2}$ & - & 244/189 \\
+ MKCFLOW) & $3.24^{+.18}_{-.17}$ & $3.65^{+.72}_{-.70}$ & $0.42^{+.07}_{-.06}$ & $4.8^{+4.4}_{-4.4}$ & - & 236/188 \\
&&&&&& \\
WABS*(MEKAL& $3.15^{+.17}_{-.15}$ & 4.88 & $0.38^{+.06}_{-.05}$ & $7.4^{+4.5}_{-3.2}$ & $<3.1$ & 244/188 \\
+ZWABS*MKCFLOW) & $3.25^{+.22}_{-.18}$ & $3.32^{+1.0}_{-2.3}$ & $0.42^{+.07}_{-.06}$ & $5.8^{+19.8}_{-5.2}$ & $<4.2$ & 236/187 \\
\hline
\end{tabular}

Notes: see the notes for Table 1.

\end{center}

\end{table*}

\begin{table*}

\begin{center}

TABLE 3

Comparison of Spectral Fits for the $29^{\prime\prime}-43^{\prime\prime}$ annulus
\begin{tabular}{c|ccccc}
\hline\hline
Model & kT & $N_H$ & Z & $\rm\dot M$ & $\chi^2$/DOF \\
& (keV) & $(10^{20} \rm{cm}^{-2})$ & (solar) & $(\Mo \rm{yr}^{-1})$ &  \\
\hline
WABS*MEKAL & $3.25^{+.12}_{-.12}$ & 4.88 & $0.34^{+.06}_{-.05}$ & - &  259/232 \\
 & $3.36^{+.14}_{-.15}$ & $4.03^{+.70}_{-.67}$ & $0.38^{+.06}_{-.06}$ & - & 255/231 \\
&&&&& \\
WABS*(MEKAL& $3.39^{+.18}_{-.17}$ & 4.88 & $0.36^{+.06}_{-.05}$ & $< 9.9$ & 256/231 \\
+ MKCFLOW) & $3.43^{+.18}_{-.17}$ & $4.22^{+.72}_{-.71}$ & $0.38^{+.07}_{-.06}$ & $<8.1$ & 254/230 \\
\hline
\end{tabular}

Notes: see the notes for Table 1.

\end{center}

\end{table*}

\newpage


\begin{figure*}[tb]
\pspicture(0,9.8)(18.5,22.9)

\rput[tl]{0}(5.0,27.2){\epsfxsize=8.5cm
\epsffile{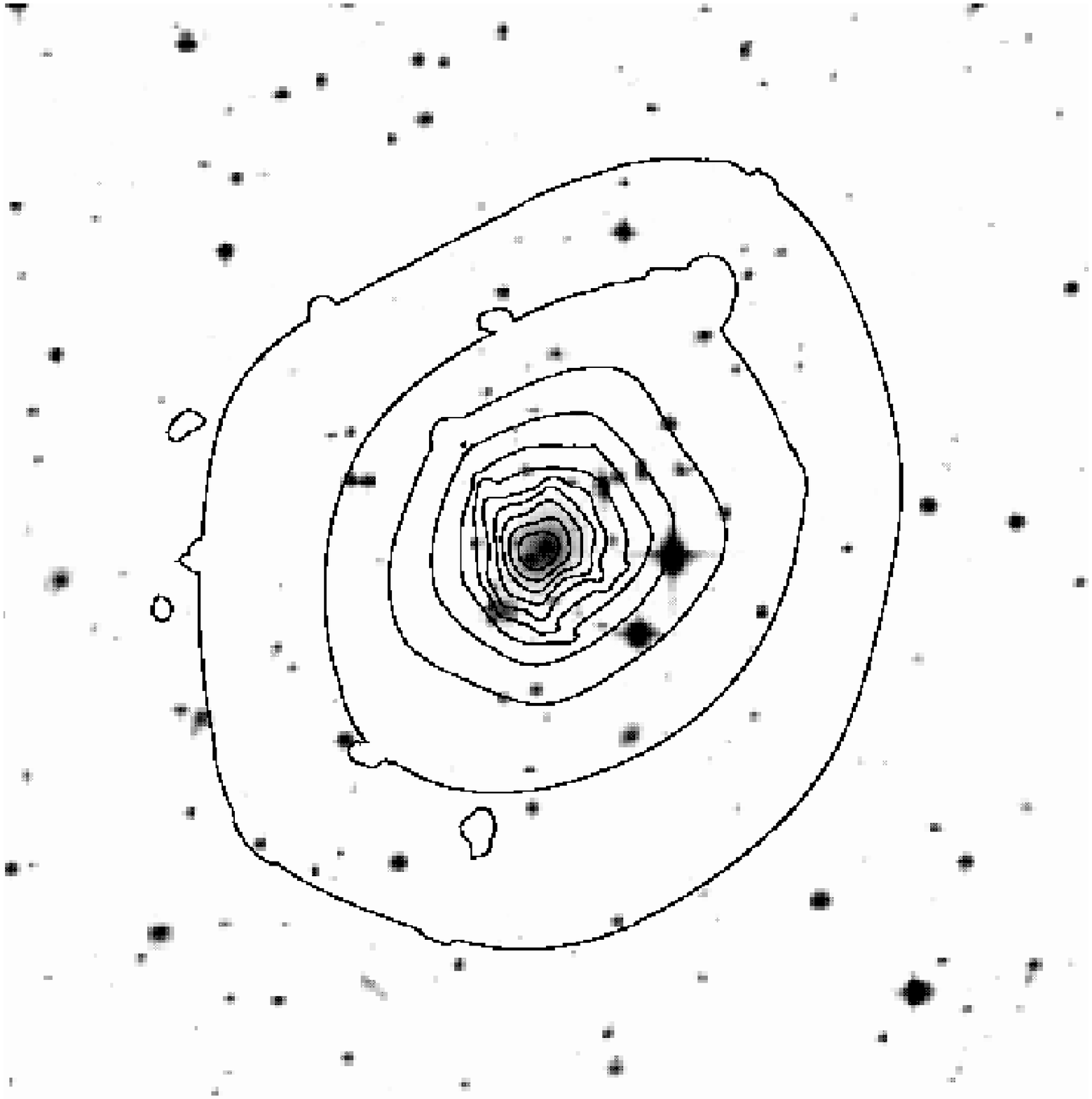}}

\rput[tl]{0}(0,18.5){
\begin{minipage}{18cm}
\small\parindent=3.5mm
{\sc Fig.}~1. ACIS-I contours of the background subtracted, vignetting and
exposure corrected ACIS-I data overlaid on the optical image of the Hydra A
cluster.  The figure is $8.53^{\prime}$ (540 kpc) on a side.
\end{minipage}
}
\rput[tl]{0}(5.0,16.2){\epsfxsize=7.75cm
\epsffile[30 470 530 678]{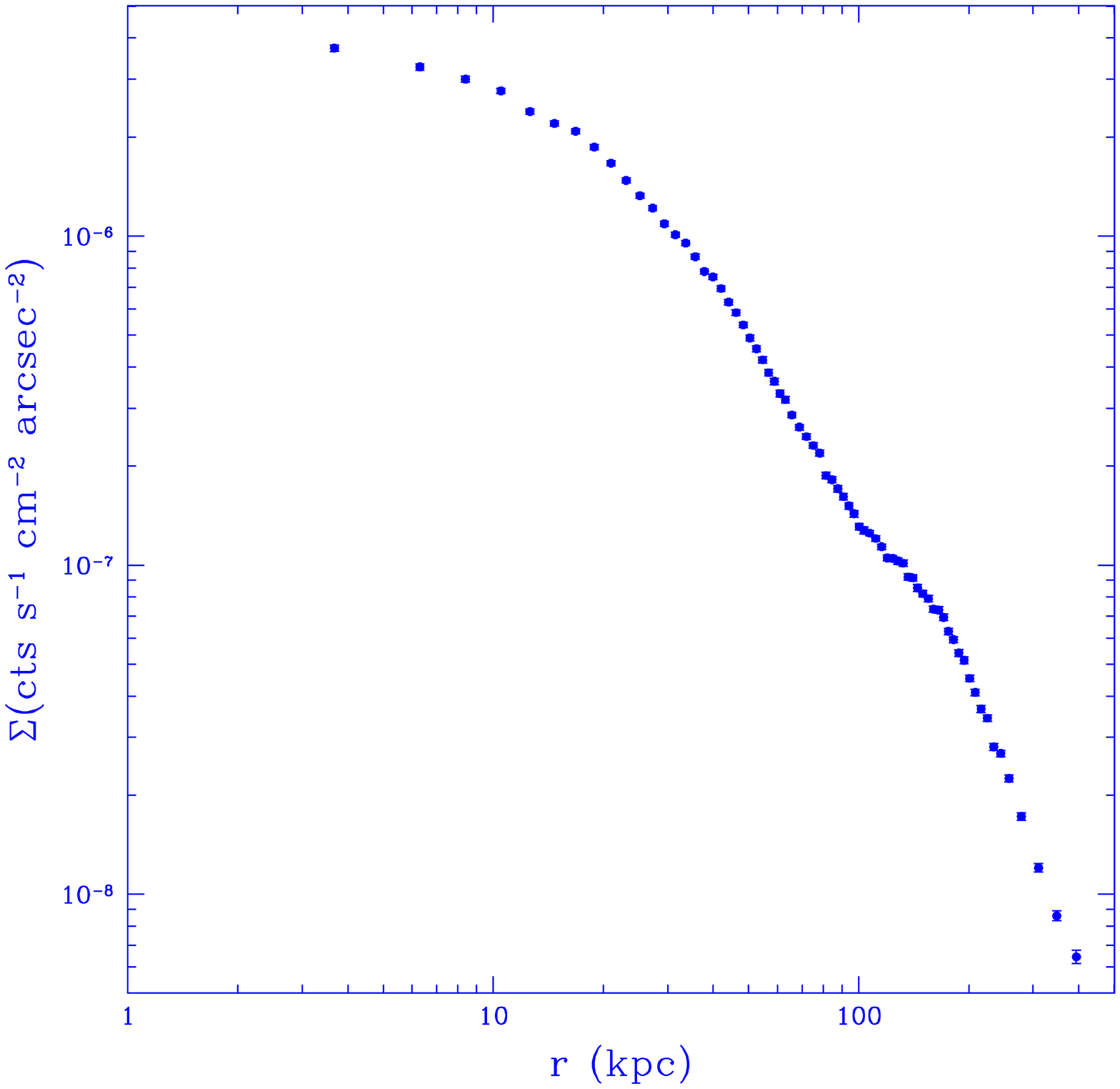}}

\rput[tl]{0}(0,8.0){
\begin{minipage}{18cm}
\small\parindent=3.5mm
{\sc Fig.}~2. The 0.3-7.0~keV surface brightness profile of the Hydra A cluster from the
ACIS-S observation.  The data are background subtracted, vignetting
and exposure corrected, and binned to produce 2000 net counts per annulus.
\end{minipage}
}
\endpspicture
\end{figure*}

\newpage


\begin{figure*}[tb]
\pspicture(0,9.8)(18.5,22.9)

\rput[tl]{0}(5.0,27.2){\epsfxsize=8.5cm
\epsffile{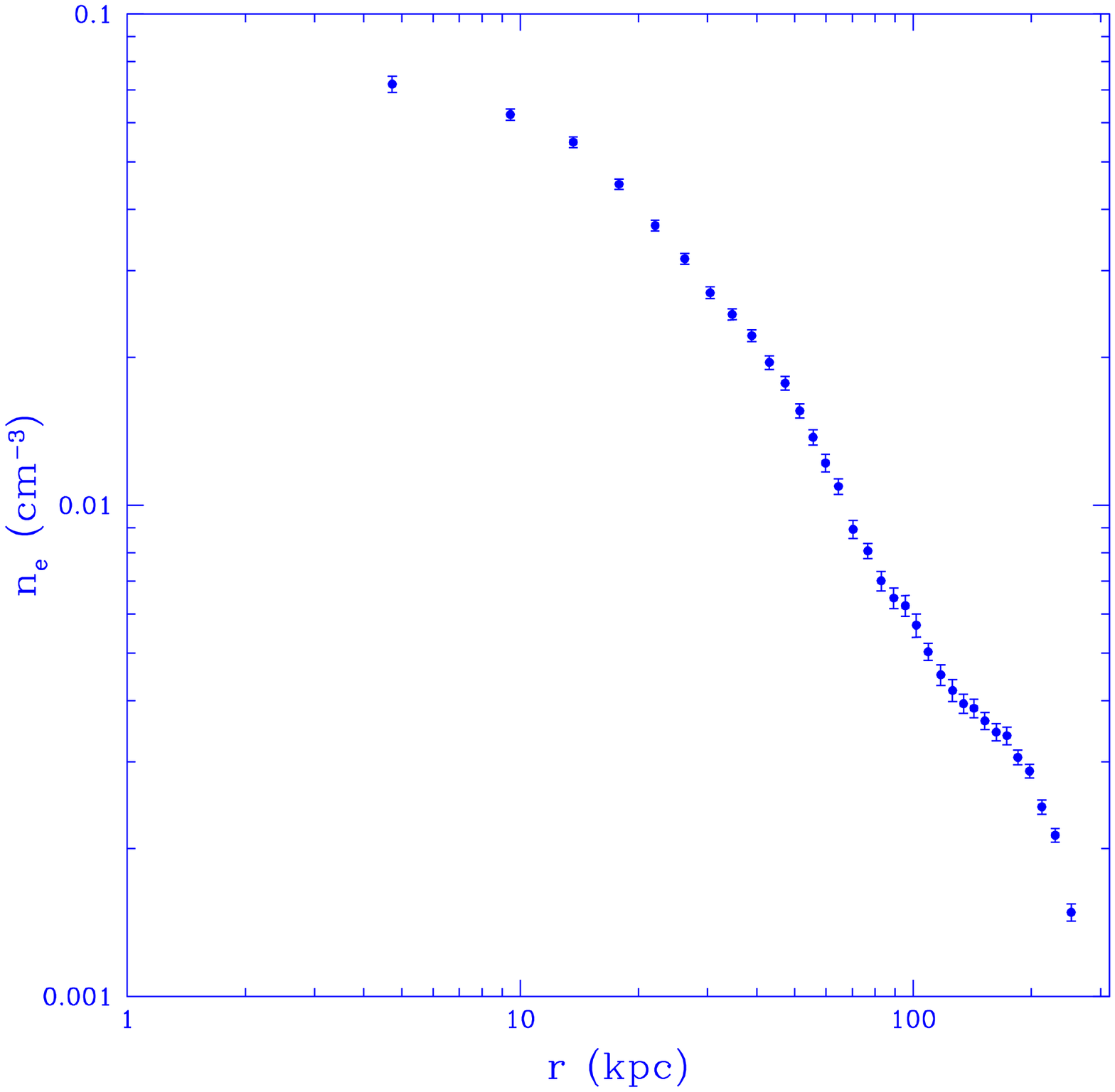}}

\rput[tl]{0}(0,18.5){
\begin{minipage}{18cm}
\small\parindent=3.5mm
{\sc Fig.}~3. The deprojected electron density profile of the Hydra A cluster with
errors derived from 100 Monte Carlo simulations.
\end{minipage}
}
\rput[tl]{0}(5.0,16.2){\epsfxsize=7.75cm
\epsffile[30 470 530 678]{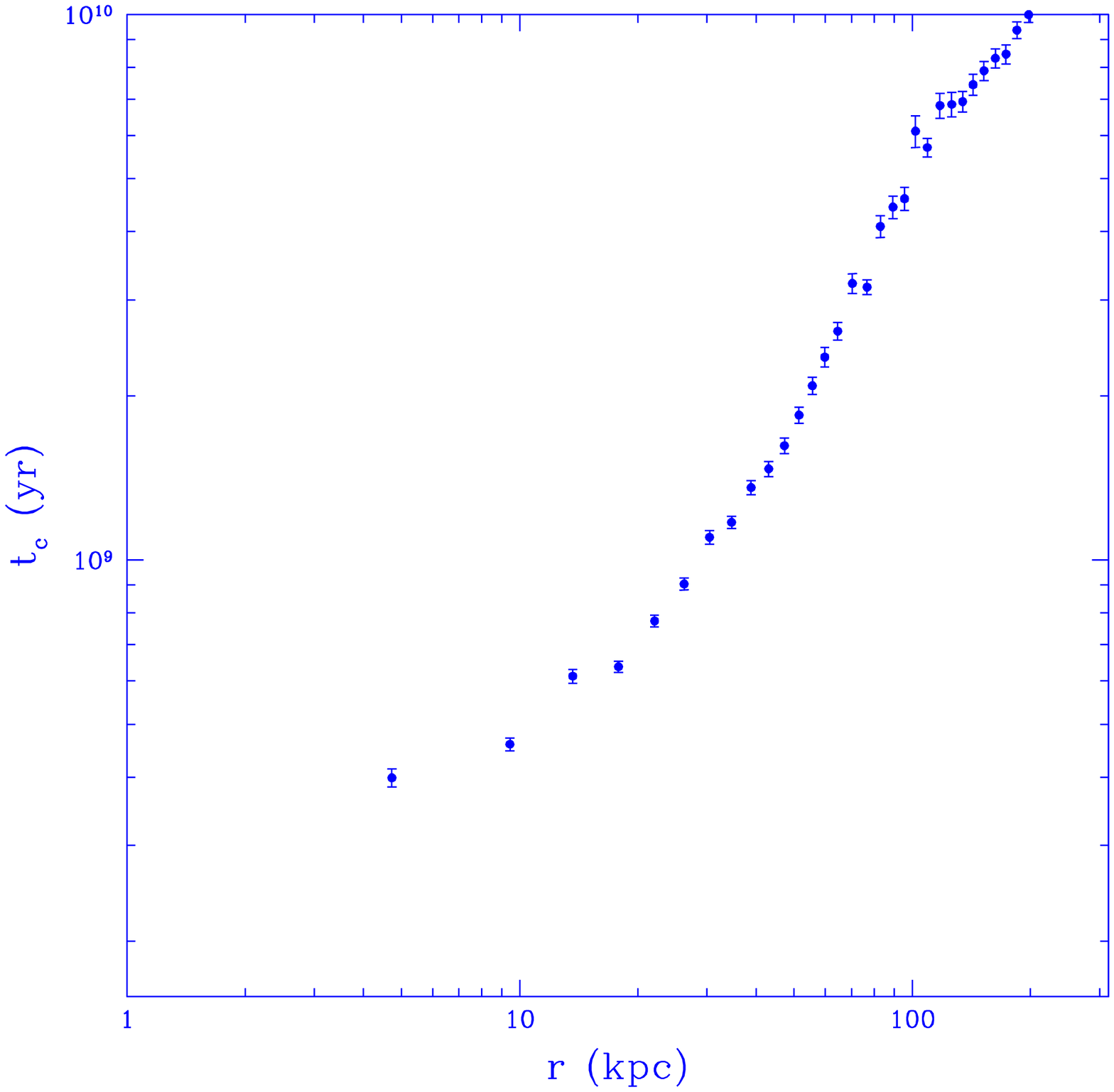}}

\rput[tl]{0}(0,8.0){
\begin{minipage}{18cm}
\small\parindent=3.5mm
{\sc Fig.}~4. The isobaric cooling time of the hot gas as a function of radius.
\end{minipage}
}

\endpspicture
\end{figure*}

\newpage


\begin{figure*}[tb]
\pspicture(0,9.8)(18.5,22.9)

\rput[tl]{0}(5.0,27.2){\epsfxsize=8.5cm
\epsffile[30 470 630 678]{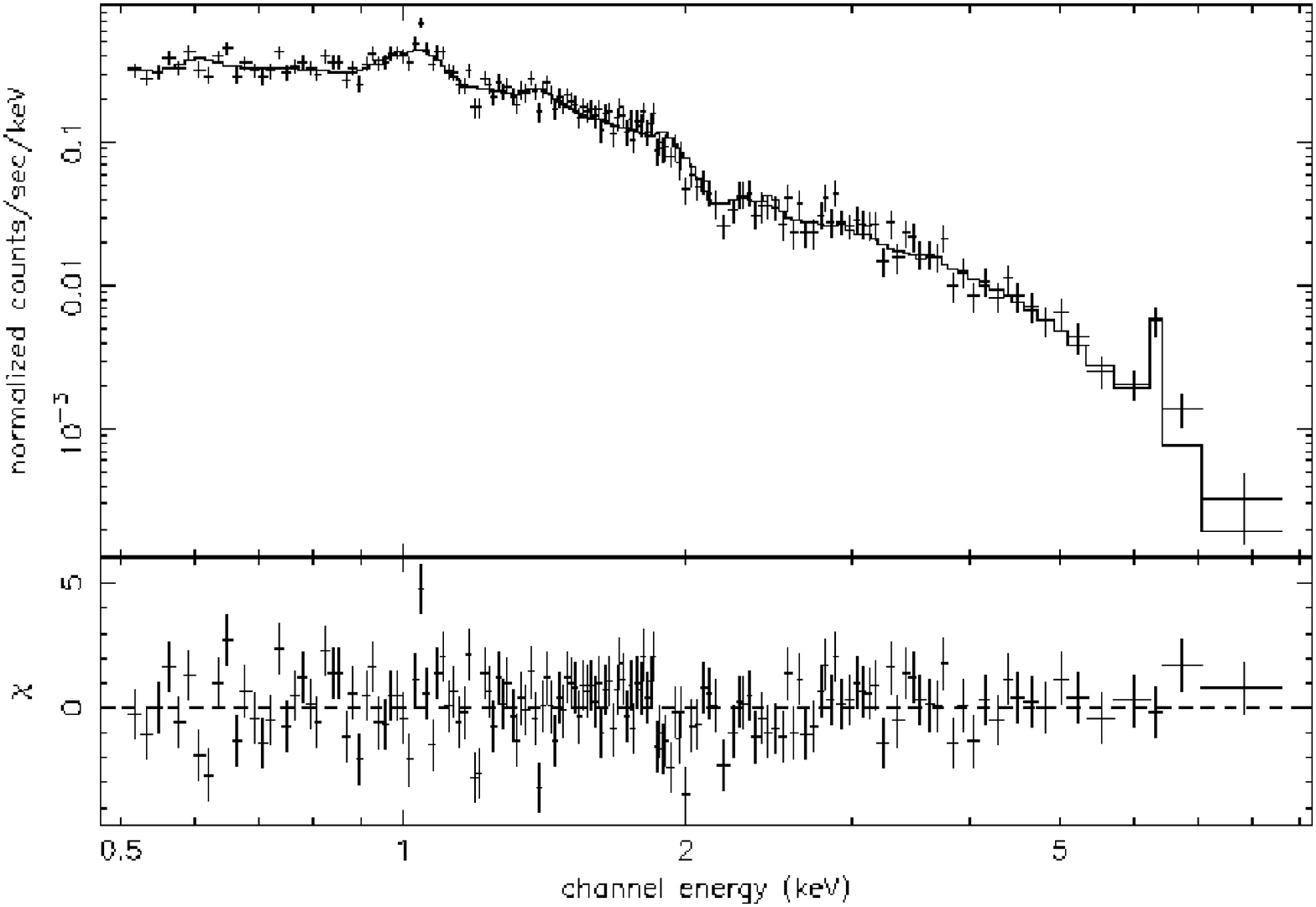}}

\rput[tl]{0}(5.6,16.2){\epsfxsize=8.5cm
\epsffile[30 470 630 678]{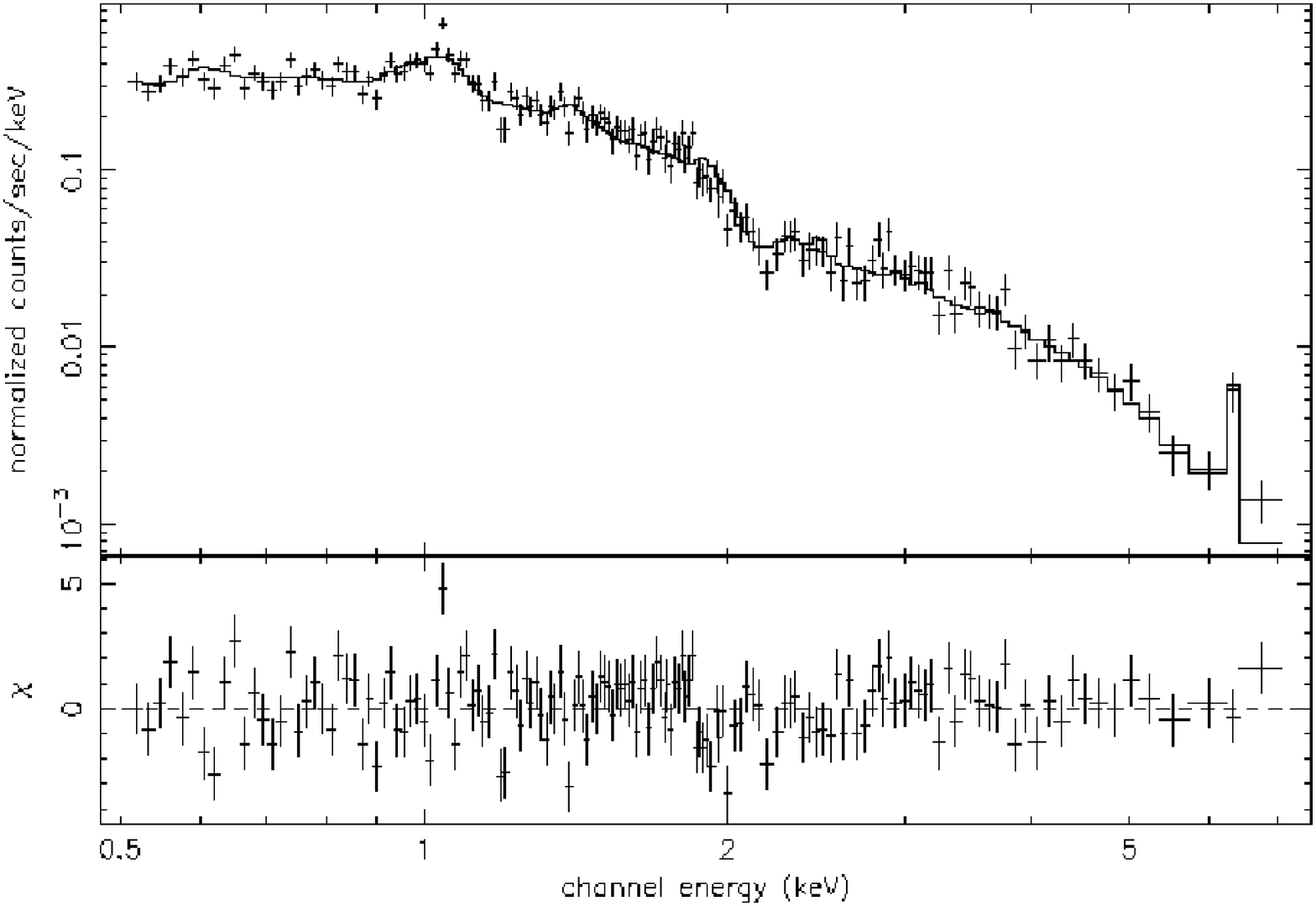}}

\rput[tl]{0}(0,8.0){
\begin{minipage}{18cm}
\small\parindent=3.5mm
{\sc Fig.}~5. a) Best-fit single temperature model to the innermost ACIS-S
spectrum ($2^{\prime\prime}$-$17^{\prime\prime}$), and b) best-fit
single temperature plus cooling flow model to the innermost ACIS-S
spectrum.
\end{minipage}
}
\endpspicture
\end{figure*}

\newpage


\begin{figure*}[tb]
\pspicture(0,9.8)(18.5,22.9)

\rput[tl]{0}(5.0,27.2){\epsfxsize=8.5cm
\epsffile[30 470 630 678]{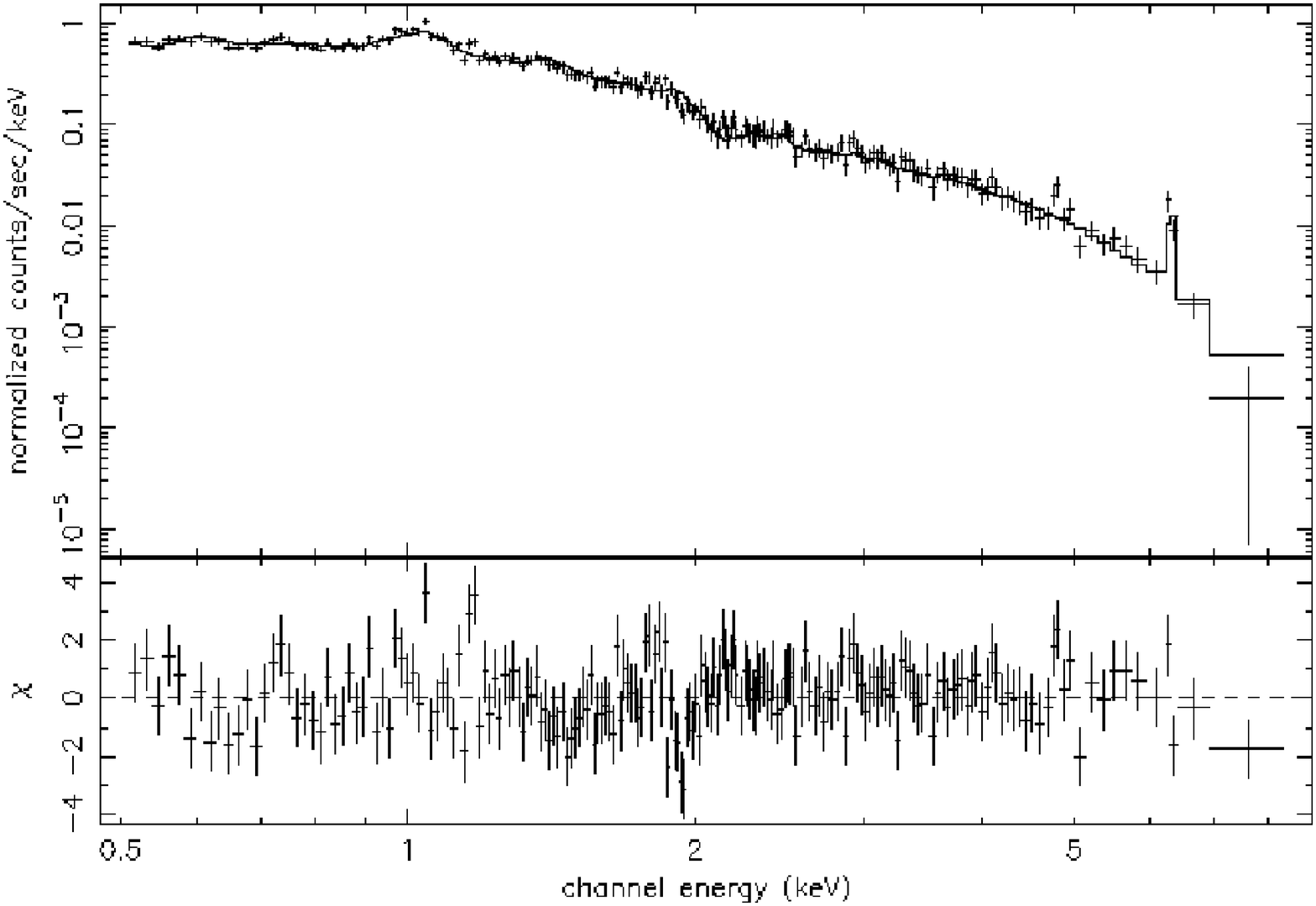}}

\rput[tl]{0}(0,18.5){
\begin{minipage}{18cm}
\small\parindent=3.5mm
{\sc Fig.}~6. Best-fit single temperature model to the ACIS-S
spectrum extracted from an annulus between $17^{\prime\prime}$
and $29^{\prime\prime}$ from the central point source.
\end{minipage}
}

\rput[tl]{0}(5.0,17.2){\epsfxsize=8.5cm
\epsffile[30 470 630 678]{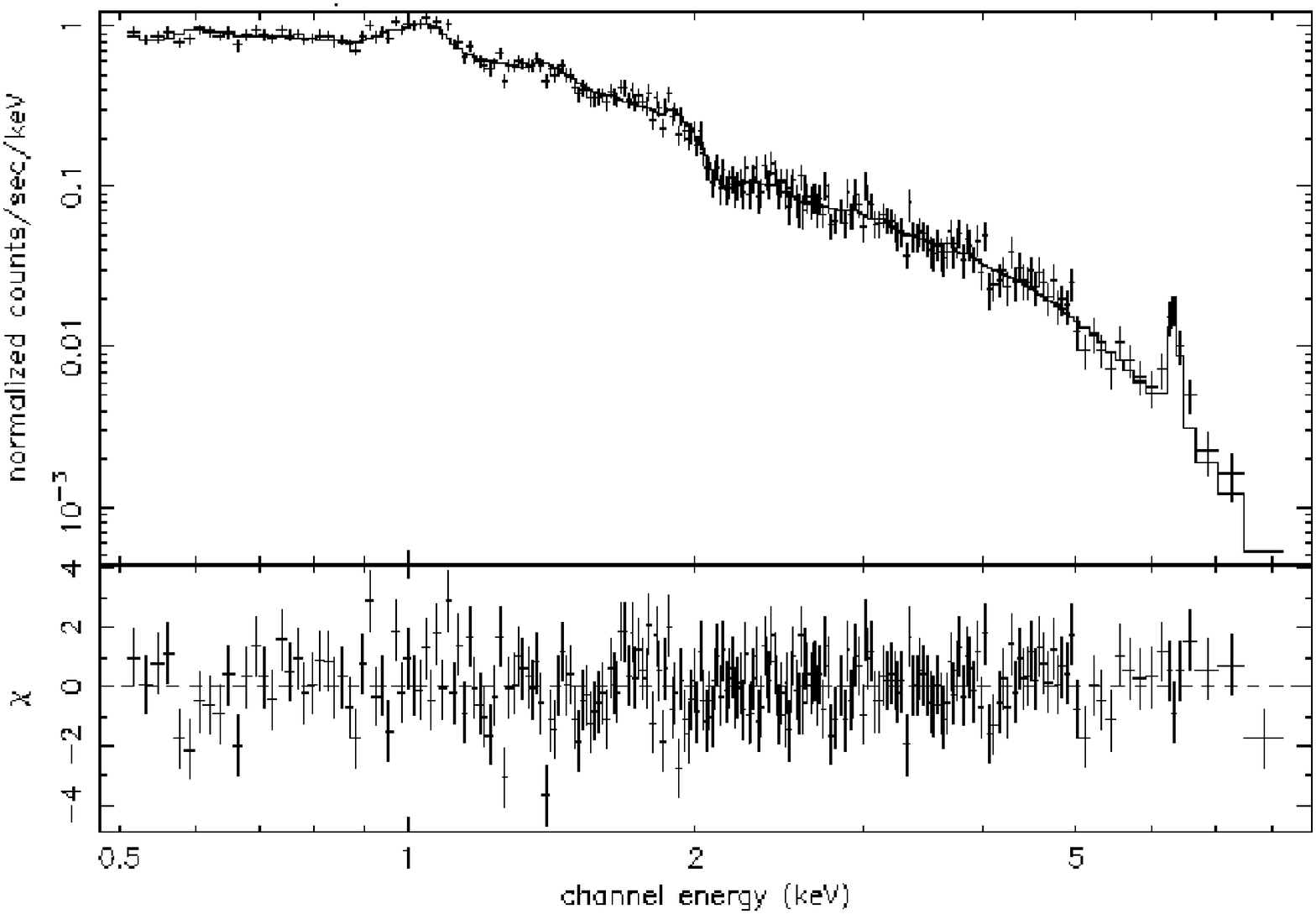}}

\rput[tl]{0}(0,8.0){
\begin{minipage}{18cm}
\small\parindent=3.5mm
{\sc Fig.}~7. Best-fit single temperature model to the ACIS-S
spectrum extracted from an annulus between $29^{\prime\prime}$
and $43^{\prime\prime}$ from the central point source.

\end{minipage}
}
\endpspicture
\end{figure*}

\newpage


\begin{figure*}[tb]
\pspicture(0,9.8)(18.5,22.9)

\rput[tl]{0}(5.0,27.2){\epsfxsize=8.5cm
\epsffile[30 470 630 678]{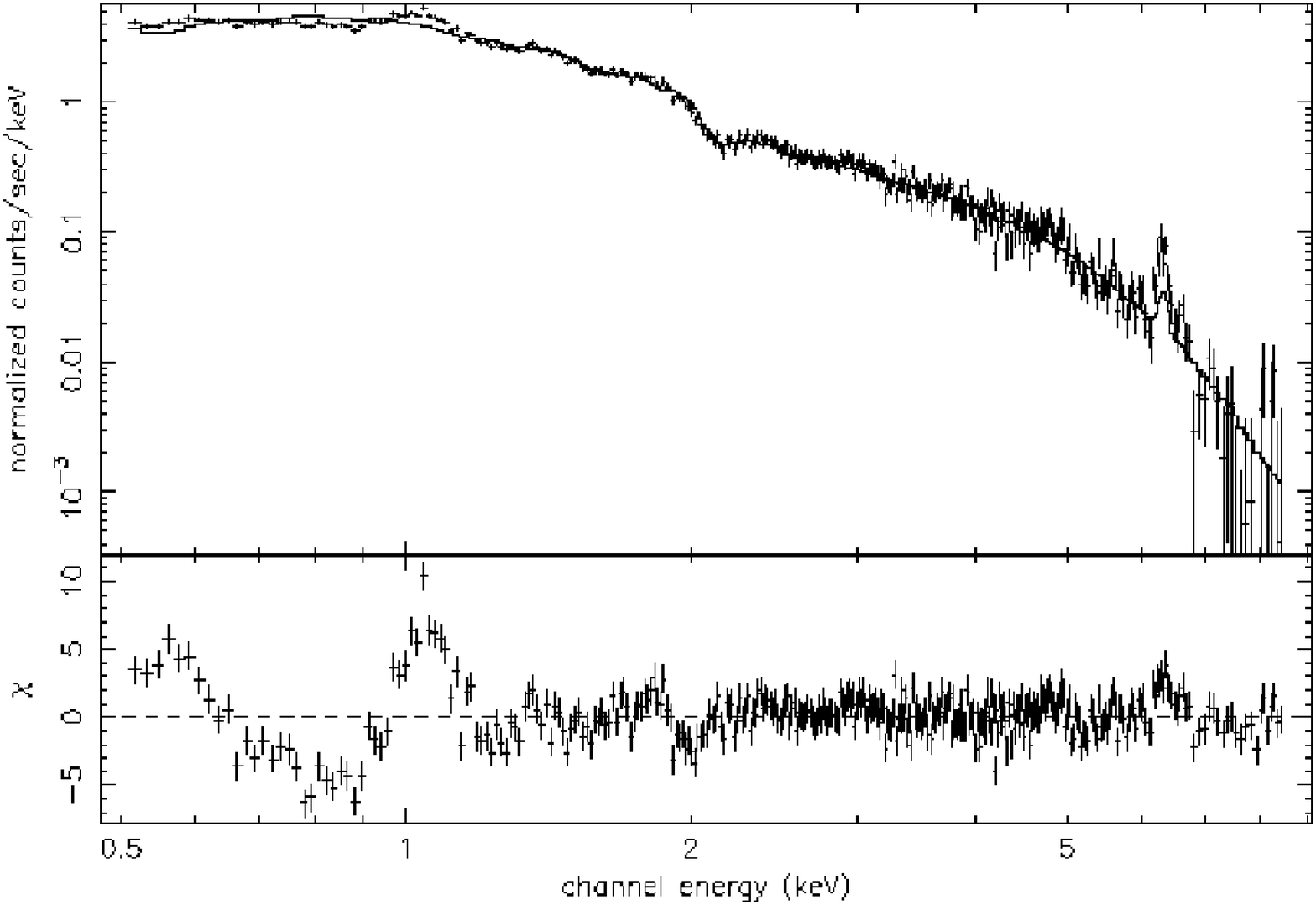}}

\rput[tl]{0}(0,18.5){
\begin{minipage}{18cm}
\small\parindent=3.5mm
{\sc Fig.}~8. Best-fit of the ACIS-S spectrum of the central 100 kpc (excluding the
central point source) to a single temperature plus cooling flow model
with $\rm\dot M$ frozen at 140~$\Mo$~yr$^{-1}$ and all other parameters free.
\end{minipage}
}

\endpspicture
\end{figure*}

\newpage


\begin{figure*}[tb]
\pspicture(0,9.8)(18.5,22.9)

\rput[tl]{0}(5.0,27.2){\epsfxsize=8.5cm
\epsffile{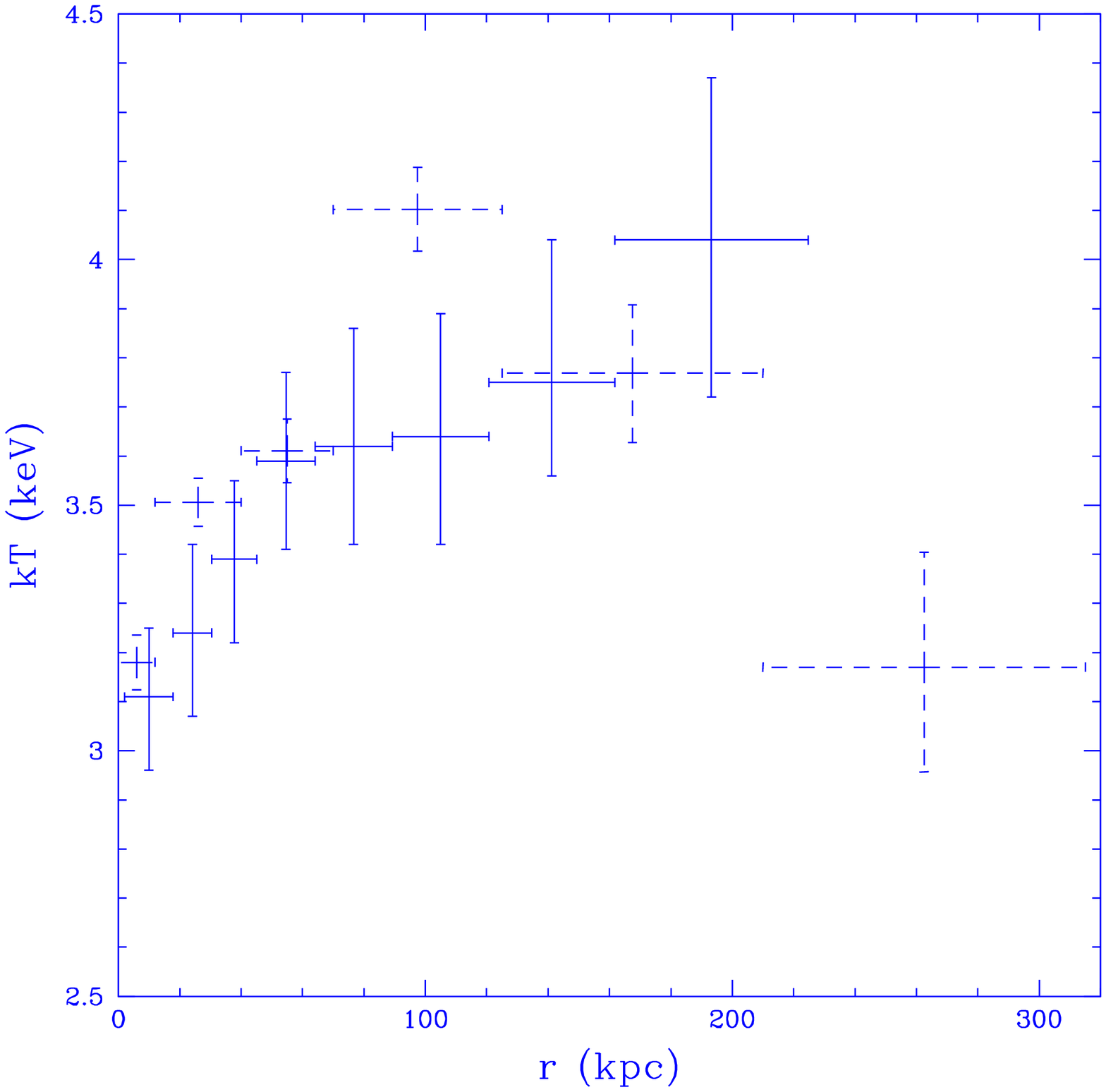}}

\rput[tl]{0}(0,18.5){
\begin{minipage}{18cm}
\small\parindent=3.5mm
{\sc Fig.}~9. Gas temperature profile of the Hydra A cluster. The solid (dashed)
crosses show the spectral analysis results of the ACIS-S (ACIS-I) data. 
Error bars are shown at the 90\% confidence level for one interesting parameter.
\end{minipage}
}
\rput[tl]{0}(5.0,16.2){\epsfxsize=7.75cm
\epsffile[30 470 530 678]{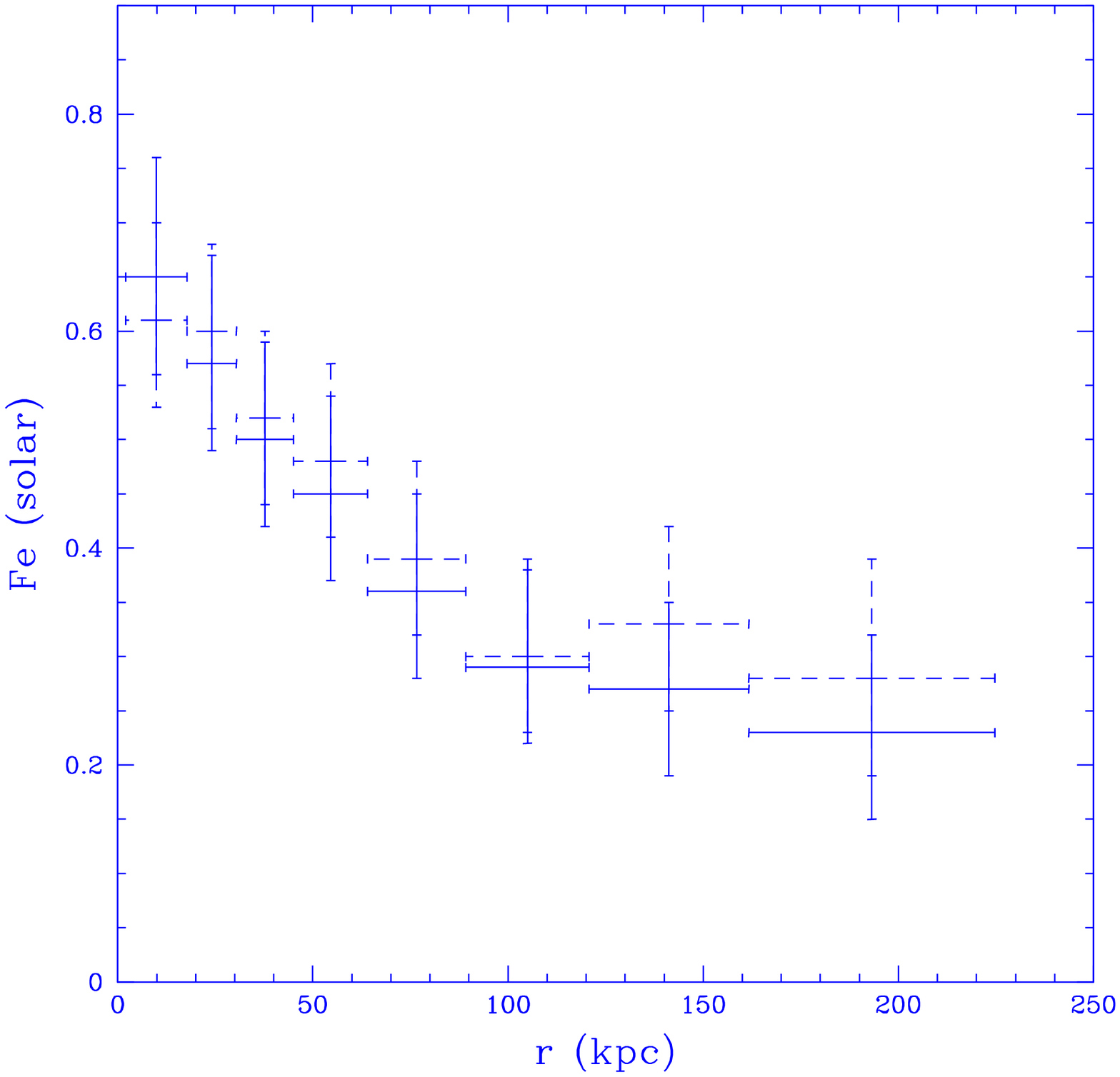}}

\rput[tl]{0}(0,8.0){
\begin{minipage}{18cm}
\small\parindent=3.5mm
{\sc Fig.}~10. Fe abundance profile relative to the solar
value ($4.68 \times 10^{-5}$ by number relative to H)
for the Hydra A cluster derived
from the ACIS-S observation. The solid lines
show the best fit to a single temperature model in the 0.5-7.5 keV band,
while the dashed lines show the the best fit to a single temperature model
in the 2.0-7.5~keV band.  Error bars are shown at the 90\% confidence
level for one interesting parameter.
\end{minipage}
}
\endpspicture
\end{figure*}

\newpage


\begin{figure*}[tb]
\pspicture(0,9.8)(18.5,22.9)

\rput[tl]{0}(5.0,27.2){\epsfxsize=8.5cm
\epsffile{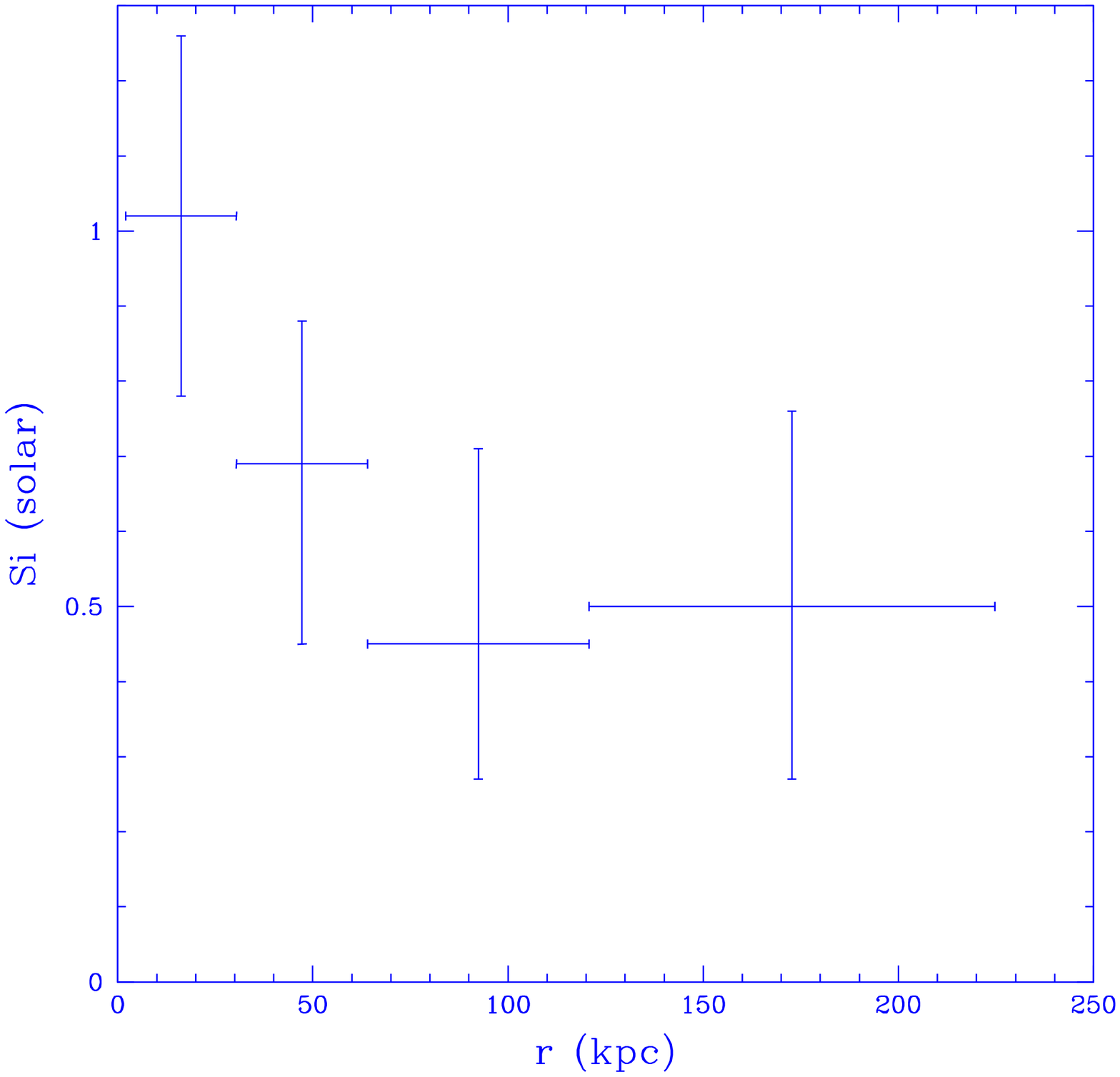}}

\rput[tl]{0}(0,18.5){
\begin{minipage}{18cm}
\small\parindent=3.5mm
{\sc Fig.}~11. Si abundance profile relative to the solar
value ($3.55 \times 10^{-5}$ by number relative to H)
for the Hydra A cluster derived
from the ACIS-S observation.  Error bars are shown at the 90\% confidence
level for one interesting parameter.
\end{minipage}
}
\rput[tl]{0}(5.0,16.2){\epsfxsize=7.75cm
\epsffile[30 470 530 678]{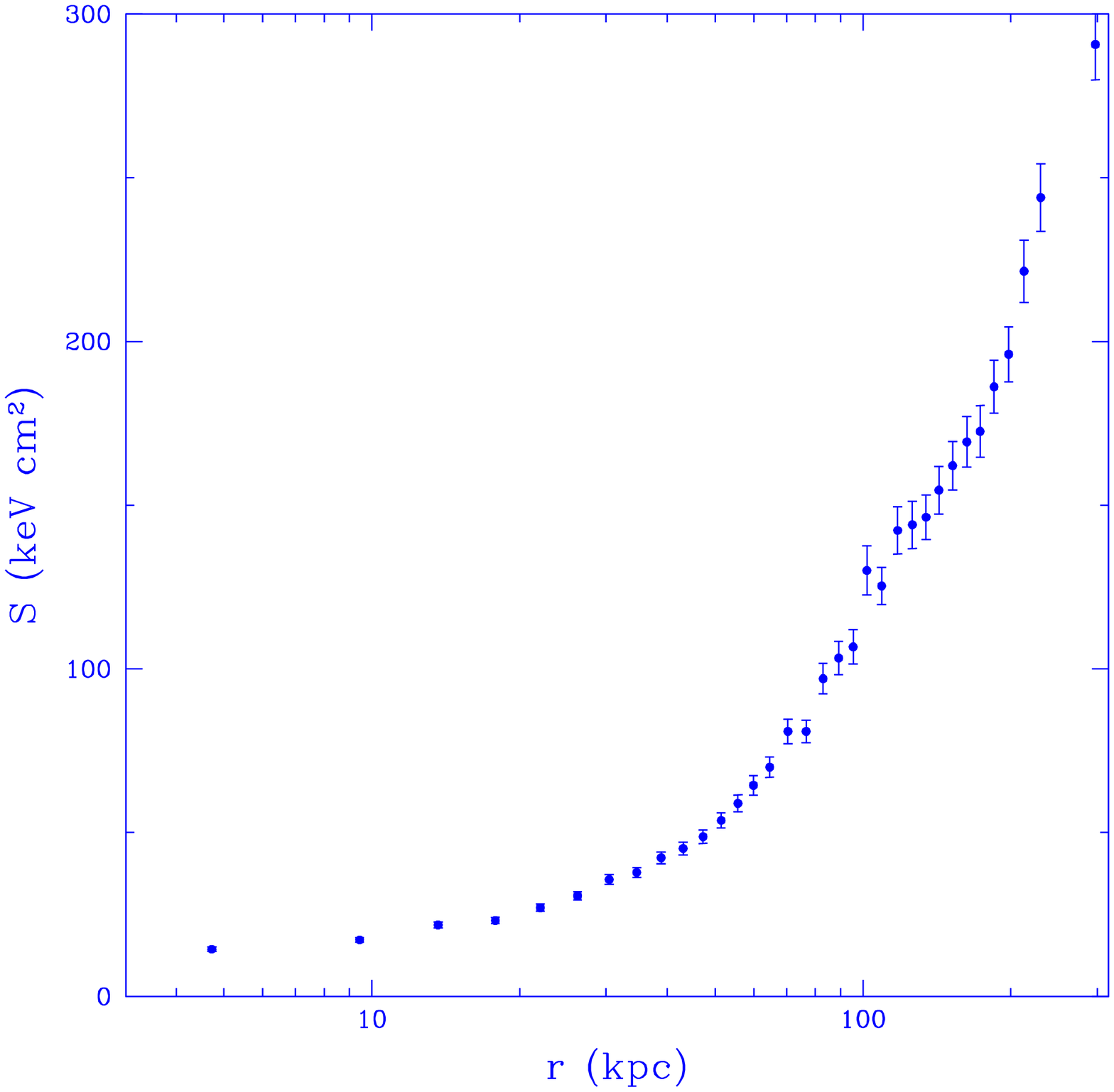}}

\rput[tl]{0}(0,8.0){
\begin{minipage}{18cm}
\small\parindent=3.5mm
{\sc Fig.}~12. Gas entropy profile derived from the deprojected density and
power-law temperature profiles.
\end{minipage}
}
\endpspicture
\end{figure*}

\newpage


\begin{figure*}[tb]
\pspicture(0,9.8)(18.5,22.9)

\rput[tl]{0}(5.0,27.2){\epsfxsize=8.5cm
\epsffile{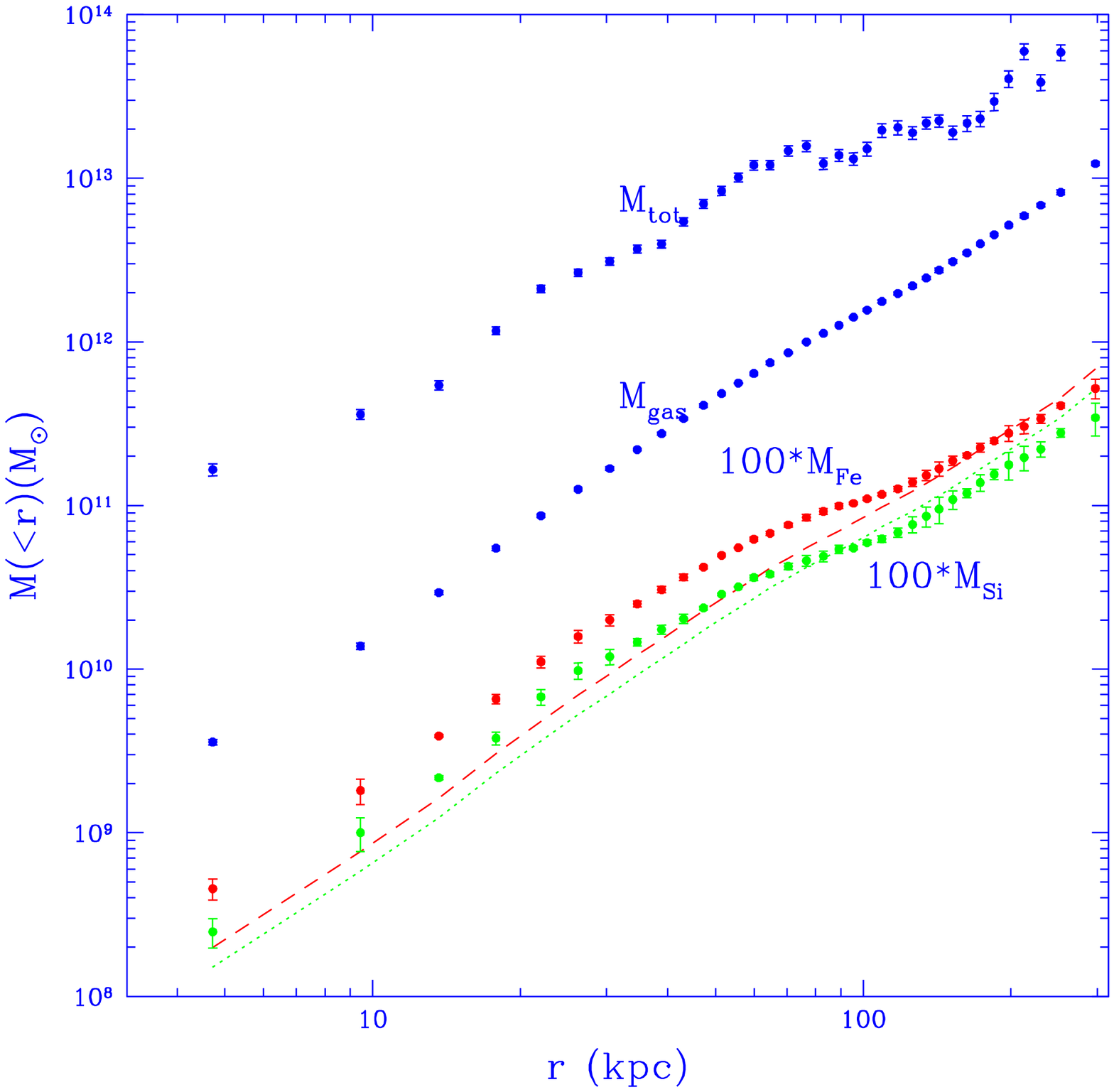}}

\rput[tl]{0}(0,18.5){
\begin{minipage}{18cm}
\small\parindent=3.5mm
{\sc Fig.}~13. The integrated total gravitating mass, gas mass, Fe mass, and 
Si mass distributions.
The Fe and Si masses are shown at 100 times their true values.  The dashed and
dotted lines are the Fe and Si masses that would be inferred assuming a uniform 
distribution of heavy elements.

\end{minipage}
}
\rput[tl]{0}(5.0,16.2){\epsfxsize=7.75cm
\epsffile[30 470 530 678]{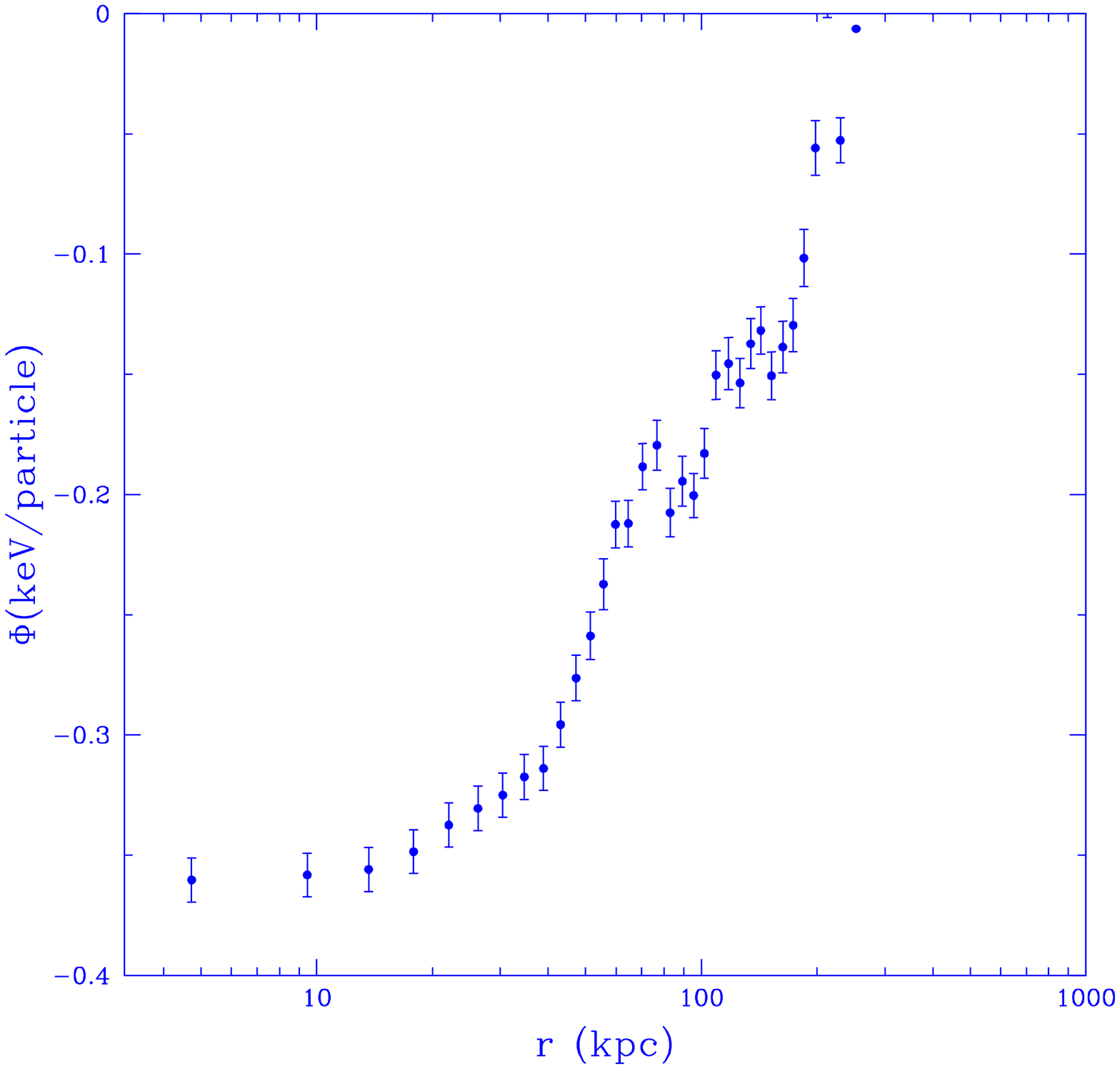}}

\rput[tl]{0}(0,8.0){
\begin{minipage}{18cm}
\small\parindent=3.5mm
{\sc Fig.}~14. The gravitational potential in the Hydra A cluster derived
from the total mass distribution shown in Figure 8.  The potential
is normalized to zero at 250 kpc.

\end{minipage}
}
\endpspicture
\end{figure*}

\newpage


\begin{figure*}[tb]
\pspicture(0,9.8)(18.5,22.9)

\rput[tl]{0}(5.0,27.2){\epsfxsize=8.5cm
\epsffile{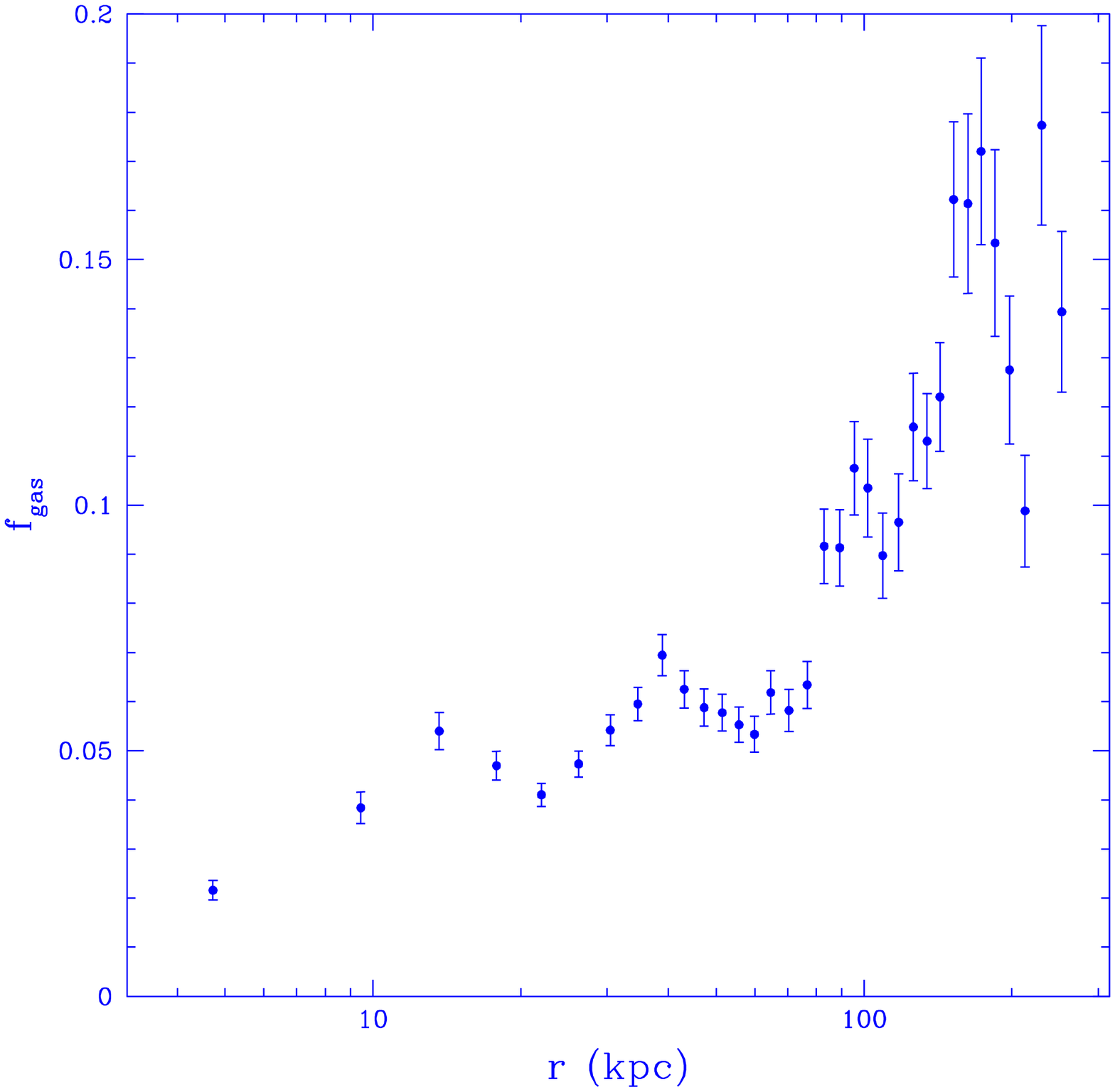}}

\rput[tl]{0}(0,18.5){
\begin{minipage}{18cm}
\small\parindent=3.5mm
{\sc Fig.}~15. Gas mass fraction versus radius in the Hydra A cluster.
\end{minipage}
}
\rput[tl]{0}(5.0,16.2){\epsfxsize=7.75cm
\epsffile[30 470 530 678]{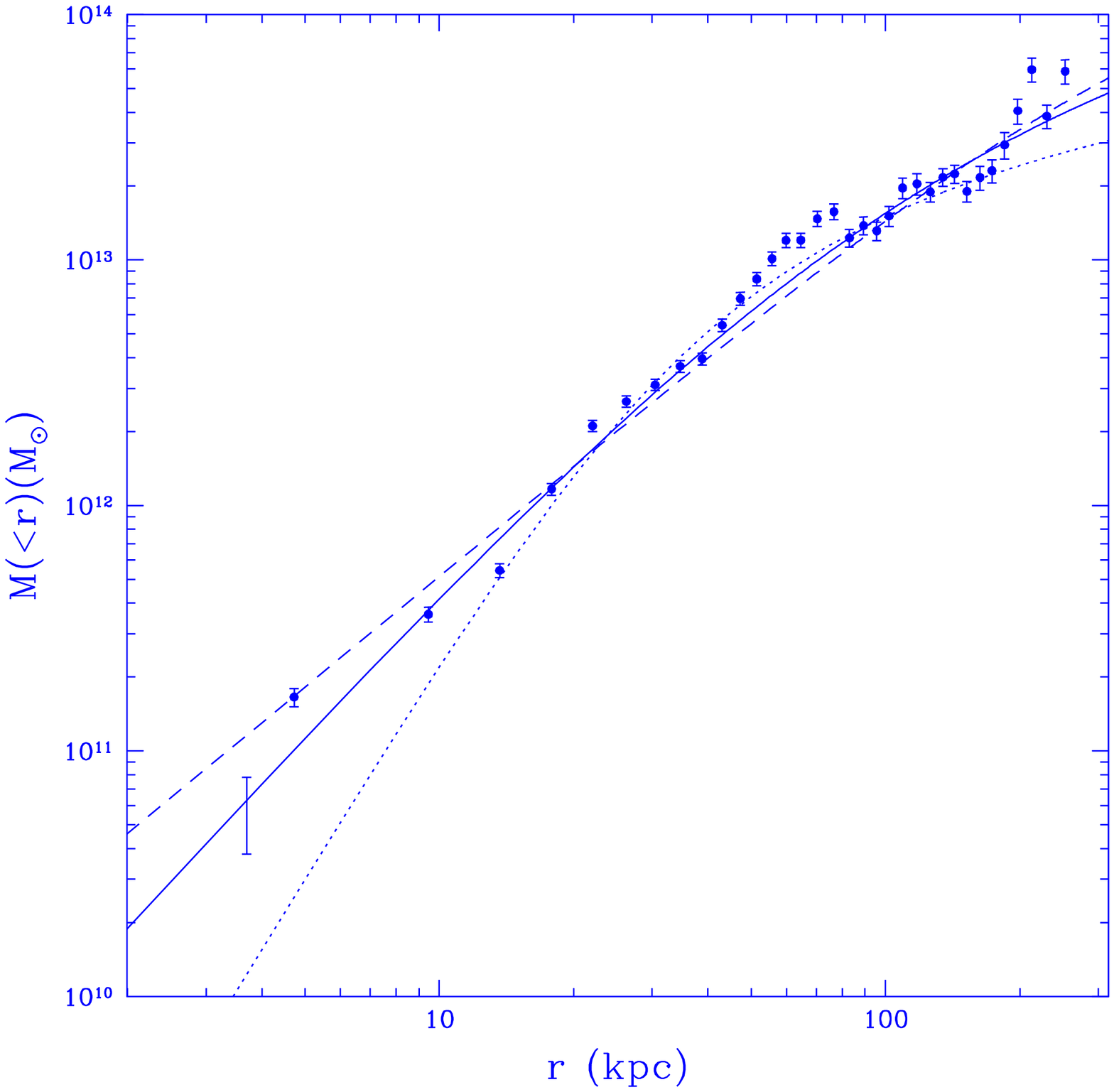}}

\rput[tl]{0}(0,8.0){
\begin{minipage}{18cm}
\small\parindent=3.5mm
{\sc Fig.}~16. Comparison of the best-fit NFW profile (solid line),
Moore profile (dashed line), and modified King profile (dotted line)
to the cumulative gravitating mass distribution in Hydra A.
The point at 3.7~kpc is the mass inferred from the 
rotation curve of the $\rm H\alpha$ disk (Melnick $\etal$ 1997).
\end{minipage}
}
\endpspicture
\end{figure*}


\begin{figure*}[tb]
\pspicture(0,9.8)(18.5,22.9)

\rput[tl]{0}(5.0,27.2){\epsfxsize=8.5cm
\epsffile{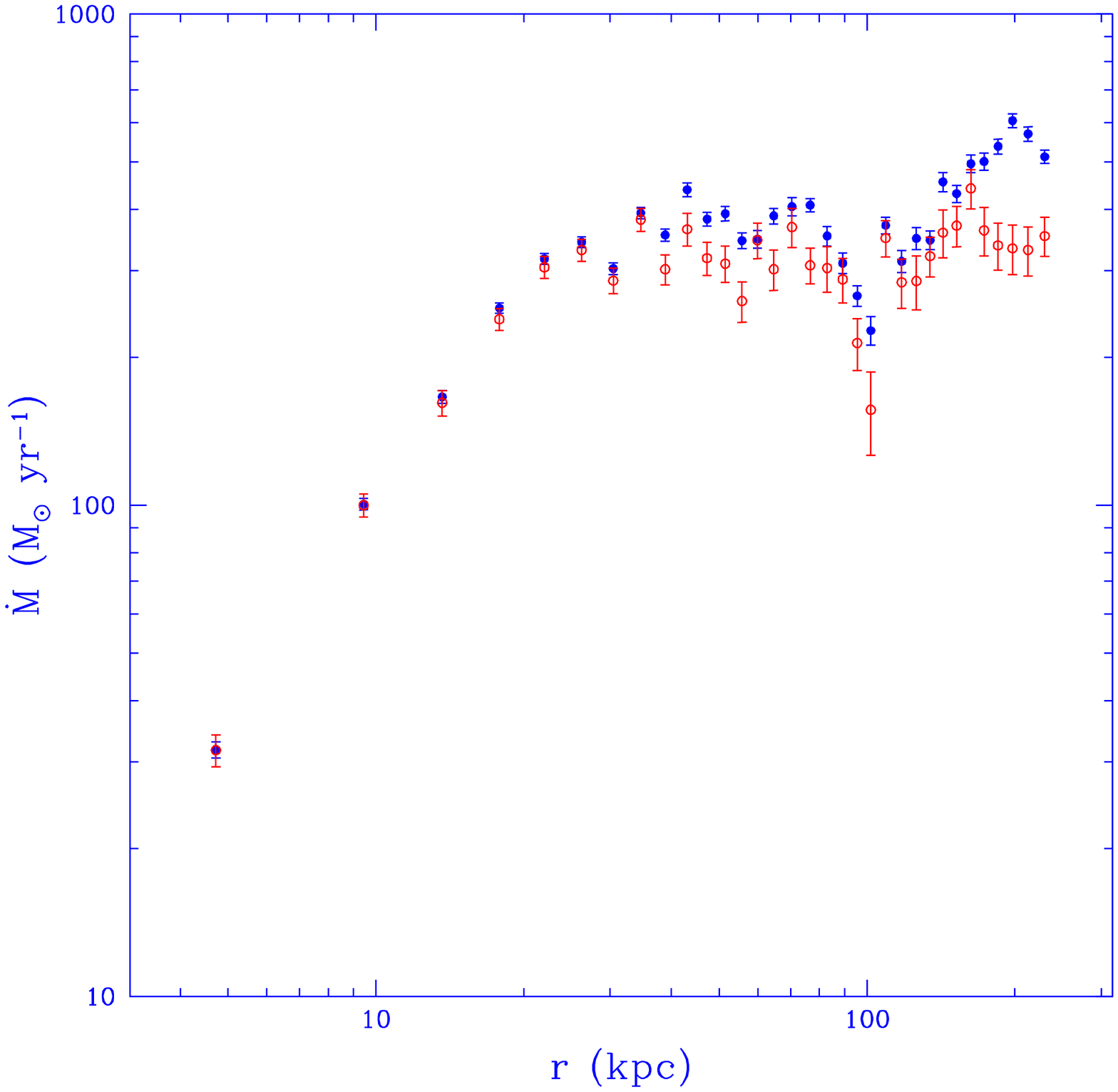}}

\rput[tl]{0}(0,18.5){
\begin{minipage}{18cm}
\small\parindent=3.5mm
{\sc Fig.}~17. Morphological $\rm\dot M$ profile derived from the deprojected
surface brightness and temperature profiles. Filled circles only include the
change in gas enthalpy across a shell, while the open circles include
the change in enthalpy and gravitational potential in eq. (4).
\end{minipage}
}
\endpspicture
\end{figure*}
\end{document}